\documentclass[12pt]{article}

\usepackage{scicite}
\usepackage{epsfig,graphicx}
\usepackage{graphics}
\usepackage{amsmath}
\usepackage{amssymb}
\usepackage{multirow}
\usepackage{tabularx, booktabs}
\newcolumntype{Y}{>{\centering\arraybackslash}X}
\usepackage{array}
\usepackage{times}

\newenvironment{sciabstract}{%
\begin{quote} \bf}
{\end{quote}}

\newcounter{lastnote}

\title{Broken translational and rotational symmetry via charge stripe order in underdoped YBa$_{2}$Cu$_3$O$_{6+y}$} 

\author{R. Comin,$^{\ast,1,2}$ R. Sutarto,$^{3}$ E. H. da Silva Neto,$^{1,2,4,5}$ L. Chauviere,$^{1,2,5}$\\R. Liang,$^{1,2}$ W. N. Hardy,$^{1,2}$ D. A. Bonn,$^{1,2}$ F. He,$^{3}$ G. A. Sawatzky,$^{1,2}$\\and A. Damascelli$^{\ast,1,2}$\\
\\
\normalsize{$^{1}$Department of Physics {\rm {\&}} Astronomy, University of British Columbia,}\\
\normalsize{Vancouver, British Columbia V6T\,1Z1, Canada}\\
\normalsize{$^{2}$Quantum Matter Institute, University of British Columbia,}\\ \normalsize{Vancouver, British Columbia V6T\,1Z4, Canada}\\
\normalsize{$^{3}$Canadian Light Source, Saskatoon, Saskatchewan S7N\,2V3, Canada}\\
\normalsize{$^{4}$Quantum Materials Program, Canadian Institute for Advanced Research,}\\ \normalsize{Toronto, Ontario M5G\,1Z8, Canada}\\
\normalsize{$^{5}$Max Planck Institute for Solid State Research,}\\
\normalsize{Heisenbergstrasse 1, D-70569 Stuttgart, Germany}\\
\\
\normalsize{$^\ast$To whom correspondence should be addressed.}\\
\normalsize{E-mail: r.comin@utoronto.ca (R.C.); damascelli@physics.ubc.ca (A.D.)}
}

\date{}

\begin{document}

\baselineskip24pt

\maketitle 

\newpage

\begin{sciabstract}
Following the early discovery of stripe-like order in La-based copper-oxide superconductors, charge ordering instabilities were observed in all cuprate families. However, it has proven difficult to distinguish between uni- (stripes) and bi-directional (checkerboard) charge order in Y- and Bi-based materials. Here we use resonant x-ray scattering (RXS) to measure the two-dimensional structure factor in YBa$_{2}$Cu$_3$O$_{6+y}$, in reciprocal space. Our data reveal the presence of charge stripe order, i.e. locally unidirectional density waves, suggesting it as the true microscopic nature of charge modulations in cuprates. At the same time, we find that the well-established competition between charge order and superconductivity is stronger for charge correlations across than along the stripes, which provides additional evidence for the intrinsic unidirectional nature of the charge order.
\end{sciabstract}

Recent studies of Y-based copper oxides have highlighted the importance of a charge-ordered electronic ground state, also termed charge-density-wave (CDW), as a central element within the phenomenology of high-temperature superconductors \cite{tranquada1995,vZimmermann_1998,abbamonte2005,hoffman2002,howald2003,hanaguri2004,wise2008,parker2010,wu2011,ghiringhelli2012,chang2012,Comin_Science,dSN_Science,Tabis_2014}. The family of YBa$_{2}$Cu$_3$O$_{6+y}$ (YBCO) compounds have yielded a wealth of experimental results that enabled advancements in our understanding of CDW instabilities and their interplay with superconductivity \cite{hinkov2004,doiron2007,Hossain,leboeuf2007,hinkov2008,daou2010,FournierNP,Ramshaw2011,wu2011,ghiringhelli2012,chang2012}.

YBCO is a layered copper-oxide-based material where hole doping is controlled by the oxygen stoichiometry in the chain layer --  characterized by uniaxial CuO chains running along the crystallographic \textbf{b} axis. In addition to ordering within the chain layer -- attained via the periodic alternation of fully-oxygenated and fully-depleted CuO chains -- recent experiments have extensively shown the presence of charge ordering in the CuO${}_{2}$ planes, with an incommensurate wavevector $Q \!\sim\! 0.31$ reciprocal lattice units \cite{Q_doping_note}, corresponding to a period of approximately 3 unit cells in real space \cite{wu2011,ghiringhelli2012,chang2012}. Although the stripy nature of La-based cuprates has been long established \cite{tranquada1995,vZimmermann_1998,abbamonte2005}, the local symmetry of the CDW in YBCO has not yet been resolved as both charge stripes (in the presence of 90$^{\circ}$ rotated domains)and checkerboard are consistent with the globally bidirectional structure of the CDW -- characterized by wavevectors along both the \textbf{a} and \textbf{b} axes, at ${\mathbf{Q}}_{a}\!\sim\!(0.31,0)$ and ${\mathbf{Q}}_{b}\!\sim\!(0,0.31)$, respectively \cite{ghiringhelli2012,chang2012,Blackburn2013,Blanco2013,LeTacon2013}. This leaves open the fundamental question as to whether stripes are the underlying charge instability in the whole class of hole-doped cuprates.

\begin{figure}[h!]
\centering
\includegraphics[width=1\linewidth]{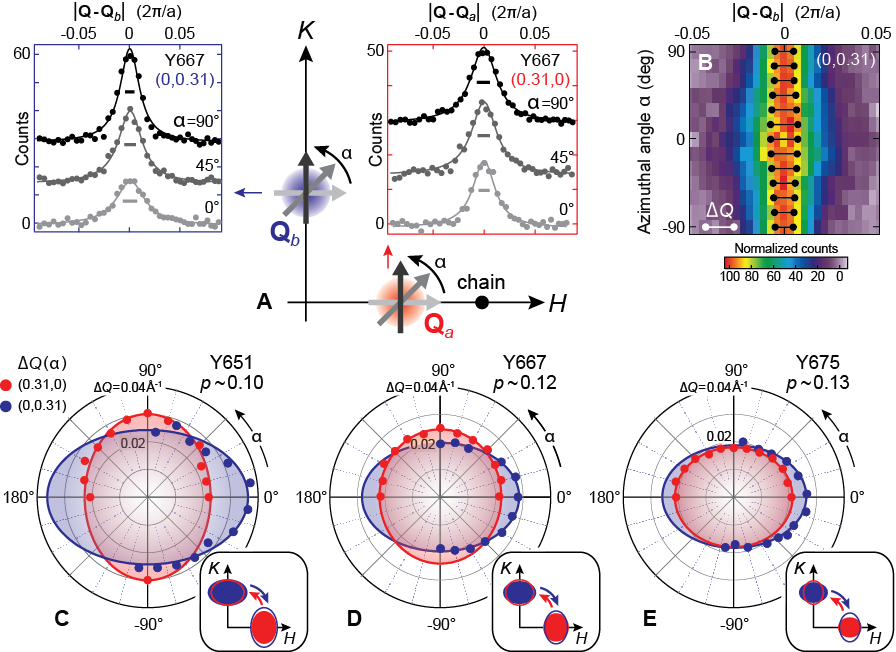}
\vspace{-2mm}
\caption{\textbf{Charge order topology in momentum space.} \textbf{(A)} Schematic representation of the momentum structure of charge modulations in YBCO. Left inset: Selected momentum scans of the CDW peak along the \textbf{b} axis at ${\mathbf{Q}}_{b}\!=\!(0,0.31)$, for different azimuthal angles ($\alpha\!=\!{0}^{\circ},{45}^{\circ},{90}^{\circ}$); continuous lines represent Lorentzian fits, horizontal bars denote the linewidth $\Delta Q$ (HWHM). Right inset: Same as for the left inset, but for the CDW peak along the \textbf{a} axis at ${\mathbf{Q}}_{a}\!=\!(0.31,0)$. \textbf{(B)} Color map of a series of Q-scans (normalized to the peak height) slicing the ${\mathbf{Q}}_{b}$ peak between $\alpha\!=\!-{90}^{\circ}$ and ${90}^{\circ}$; black bars represent the linewidth $\Delta Q$, which is largest at  $\alpha\!=\!{0}^{\circ}$. \textbf{(C-E)} Polar plots of $\Delta Q$ as a function of the azimuthal angle $\alpha$ for ${\mathbf{Q}}_{a}$ (red) and ${\mathbf{Q}}_{b}$ (blue) in YBa${}_{2}$Cu${}_{3}$O${}_{6.51}$ (Y651), YBa${}_{2}$Cu${}_{3}$O${}_{6.67}$ (Y667), and YBa${}_{2}$Cu${}_{3}$O${}_{6.75}$ (Y675), respectively. Concentric grey circles are spaced by $0.01 {\AA}^{-1}$; continuous lines are fits to an elliptic profile. Bottom-right insets: CDW peaks represented as filled ellipses and compared with their rotated versions (hollow ellipses), for each doping.}
\label{Fig1}
\end{figure}

Here we study the local density correlations of the charge-ordered state, and the interaction of the latter with superconductivity (SC) in underdoped YBCO, using resonant x-ray scattering (RXS). This technique, which is now at full maturity, represents a unique combination of diffraction (to probe reciprocal space) and resonant absorption (allowing element-specificity and therefore site-selectivity), and directly measures the structure factor $S ({Q}_{x},{Q}_{y})$, where ${Q}_{x}$ and ${Q}_{y}$ represent the momenta along the reciprocal axes H and K, respectively. The structure factor is linked to the density-density correlation function, and therefore to the CDW order parameter in momentum space (see Materials and Methods in Ref.\,\citen{SM_note}). Here, in order to reconstruct the two-dimensional (2D) structure factor with high resolution, we utilize a specifically devised RXS probing scheme whereby a charge-ordering peak is sliced along different directions, parametrized by the azimuthal angle $\alpha$ (Fig.\,\ref{Fig1}). The resulting 2D shape of the CDW peaks rules out checkerboard order and is instead consistent with a stripy nature of charge modulations in YBCO \cite{Charge_note}. We carry out RXS measurements around the CDW wavevectors ${\mathbf{Q}}_{a}\!\sim\!(0.31,0)$ and ${\mathbf{Q}}_{b}\!\sim\!(0,0.31)$ for three detwinned, oxygen-ordered YBCO samples -- YBa$_{2}$Cu$_3$O$_{6.51}$ (Y651, with hole doping $p \!\sim\! 0.10$), YBa$_{2}$Cu$_3$O$_{6.67}$ (Y667, $p \!\sim\! 0.12$), YBa$_{2}$Cu$_3$O$_{6.75}$ (Y675, $p \!\sim\! 0.13$). In our experimental scheme the CDW peaks are scanned in a radial geometry via control of the azimuthal angle $\alpha$ \cite{Comin_bond_order} (Fig.\,\ref{Fig1}A). At the Cu-${L}_{3}$ edge, the measured signal is mainly sensitive to periodic variations in the Cu-$2p \rightarrow 3d$ transition energy \cite{achkar2012,achkar2013}, i.e. to a scalar quantity, even though the detailed contribution of a pure charge modulation vs. ionic displacements to the RXS signal cannot be decoupled \cite{Displacement_note}. In addition, the poor coherence of the CDW across the CuO${}_{2}$ planes \cite{chang2012,Blackburn2013,LeTacon2013} qualifies this electronic ordering as an exquisite two-dimensional phenomenon, thus motivating our focus on the structure factor in the $({Q}_{x},{Q}_{y})$ plane. Representative scans of the CDW peak for different $\alpha$ values and at the superconducting critical temperature $T \!\sim\! {T}_{\mathrm{c}}$ are shown in the insets to Fig.\,\ref{Fig1},A for the ${\mathbf{Q}}_{b}$ and ${\mathbf{Q}}_{a}$ CDW peaks of a Y667 sample, respectively. A change in the peak half-width-at-half-maximum (HWHM) $\Delta Q$ between $\alpha\!=\!{0}^{\circ}$ and ${90}^{\circ}$ is already apparent, but is even better visualized in the color map of Fig.\,\ref{Fig1}B, which shows the sequence of Q-scans vs. azimuthal angle and the corresponding variation of $\Delta Q$ for ${\mathbf{Q}}_{b}$ in the range $\alpha \!=\! {-90}^{\circ}$ to ${90}^{\circ}$. This same procedure is applied to all three YBCO doping levels, for both the ${\mathbf{Q}}_{a}$ and ${\mathbf{Q}}_{b}$ CDW peaks; polar plots of $\Delta Q$ vs. $\alpha$ are shown in Figs.\,\ref{Fig1},C-E for Y651, Y667 and Y675, respectively. With the aid of the ellipse fits to the CDW profiles (continuous lines), four key aspects of these data stand out: (i) all peaks show a clear anisotropy between the two perpendicular directions $\alpha \!=\! {0}^{\circ}$ and ${90}^{\circ}$; (ii) for each doping, the ${\mathbf{Q}}_{a}$ and ${\mathbf{Q}}_{b}$ peaks have different shapes, and in the case of Y651 and Y667 this is even more evident as the peaks are elongated along two different directions; (iii) the departure from an isotropic case, quantified by the elongation of the CDW ellipsoids, increases towards the underdoped regime and is opposite to the evolution of orthorhombicity, which is instead maximized at optimal doping (see Supplementary Materials and Fig.\,S4 for a more detailed discussion); (iv) the peak elongation at ${\mathbf{Q}}_{a}$ and ${\mathbf{Q}}_{b}$, evolving from biaxial (Y651 and Y667) to uniaxial (Y675), is inconsistent with the doping-independence of the uniaxial symmetry of the CuO chain layer, which rules out the possibility that the observed CDW peak structure is exclusively controlled by the crystal's orthorhombic structure -- however, the uniaxial anisotropy observed for Y675 might reflect a more pronounced interaction between the Cu-O planes and chains, possibly due to the increase in orthorhombicity upon hole doping (see more detailed discussion in the Supplementary Materials).

\begin{figure}[h!]
\centering
\includegraphics[width=1\linewidth]{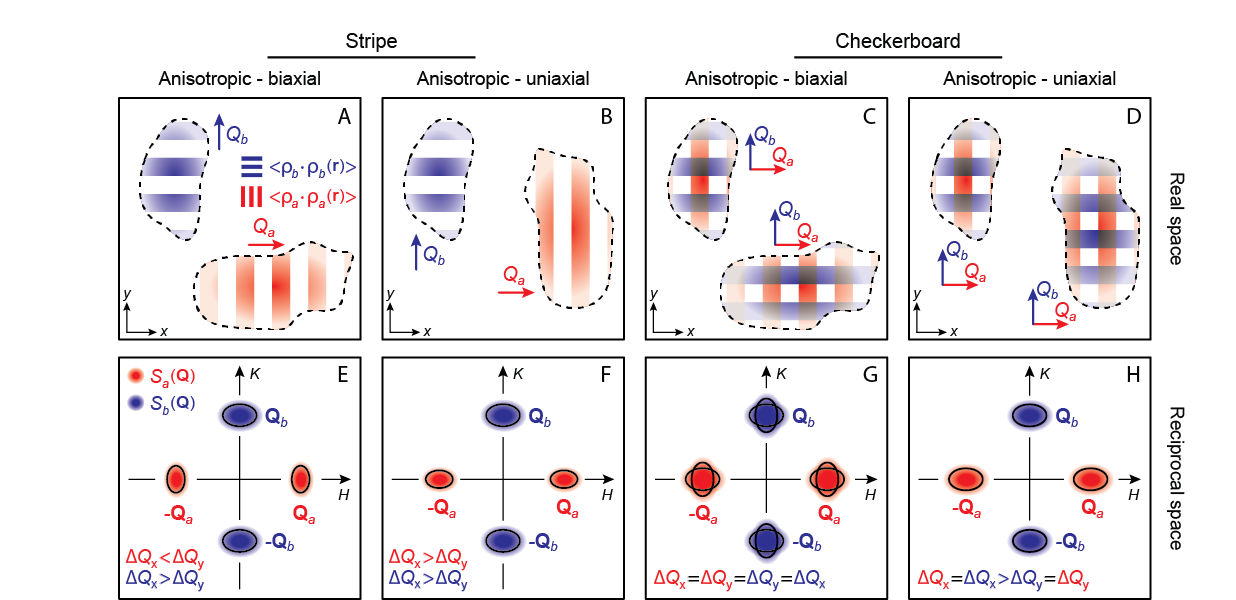}
\caption{\textbf{Domain mesostructure in real and reciprocal space.} \textbf{(A,B)} Stripy domains along \textbf{a} (red stripes) and \textbf{b} (blue stripes), in the presence of biaxial (A) and uniaxial (B) anisotropic correlations. The domains are visualized in the same layer, although a situation in which they are present in alternating layers in a ${90}^{\circ}$-rotated arrangement is equally possible. \textbf{(C,D)} Checkerboard domains in the presence of biaxial (C) and uniaxial (D) anisotropic correlations. \textbf{(E-H)} Corresponding structure factors in reciprocal (\textbf{Q}) space. In case of the simultaneous presence of both CDW components (checkerboard), the imprinted correlations must be equal for the density wave along \textbf{a} and \textbf{b}, thus imposing an equivalent peak structure at ${\mathbf{Q}}_{a}$ and ${\mathbf{Q}}_{b}$ (panels G and H).}
\label{Fig2b}
\end{figure}

The observed 2D peak structure directly bears the signatures of a breaking of fourfold (${C}_{4}$) symmetry \textit{at both the macro- and nano-scale}, and therefore of the emergence of a stripe-ordered state. In fact, under the fourfold symmetry the electronic density would be invariant under a ${90}^{\circ}$ rotation in real space ($x \rightarrow y$, $y \rightarrow -x$), which is equivalent to a ${90}^{\circ}$ rotation in momentum space (${Q}_{x} \rightarrow {Q}_{y}$, ${Q}_{y} \rightarrow -{Q}_{x}$). Instead, the CDW structure factor $S ({Q}_{x},{Q}_{y})$ is clearly not invariant under such operation, as shown in the bottom-right diagrams in Figs.\,\ref{Fig1}, C-E which compare the original $S ({Q}_{x},{Q}_{y})$ to their ${90}^{\circ}$ rotated version $S ({Q}_{y},-{Q}_{x})$. This finding demonstrates an unambiguous breaking of \textit{global} ${C}_{4}$ symmetry in all investigated samples, and might elucidate the origin of the anisotropy observed in the Nernst effect \cite{daou2010} and in optical birifrengence measurements \cite{Lubashevsky_2014}. The real-space representation of charge order branches off into two possible scenarios: (i) a \textit{biaxial anisotropy}, where both x- and y-elongated domains \cite{Domain_note} are present (Figs.\,\ref{Fig2b}, A and C); (ii) a \textit{uniaxial anisotropy}, where only y-elongated (or, equivalently, x-elongated) domains are found (Figs.\,\ref{Fig2b}, B and D). Note that these domains need not necessarily lie in the very same layer, but they need to be present at the same time within the bulk of the material (e.g., they can be present in alternating layers, while still leading to the same momentum structure). The momentum-space representation of the order parameter -- and therefore of the electronic density fluctuations -- is shown in the corresponding panels in Figs.\,\ref{Fig2b}, E-H. Here $ {S}_{a} (\mathbf{Q})$ (red ellipses) and $ {S}_{b} (\mathbf{Q})$ (blue ellipses) represent the structure factor associated to charge modulations along \textbf{a} and \textbf{b}, respectively. The profile of a single structure-factor peak is compounded with two contributions: the underlying CDW symmetry as well as its 2D correlation length, which can also be anisotropic. As a net result, the anisotropy of a single peak in Q-space cannot be used to discriminate between different CDW symmetries. Instead the latter can be resolved by probing the 2D CDW structure factor, i.e. by comparing the CDW peak shape for both ${\mathbf{Q}}_{a}$ and ${\mathbf{Q}}_{b}$. By inspecting the diagrams in Figs.\,\ref{Fig2b}, E-H, one can recognize a common trait of checkerboard structures in momentum space, in that the following conditions (Figs.\,\ref{Fig2b}, G and H) must \textit{always} be verified by symmetry: ${\Delta Q}^{a}_{x} \!=\! {\Delta Q}^{b}_{x}$ and ${\Delta Q}^{a}_{y} \!=\! {\Delta Q}^{b}_{y}$ -- i.e. \textit{the peak broadening along a given direction must coincide for} ${\Delta}_{a}$ \textit{and} ${\Delta}_{b}$ (see bottom of Figs.\,\ref{Fig2b}, E-H for case-specific conditions on the peak linewidths). Intuitively, this follows from the fact that -- for the checkerboard case -- the charge modulations along \textbf{a} and \textbf{b} axes must be subject to the same correlation lengths within each domain -- irrespective of its orientation -- and therefore lead to an equivalent peak broadening along the same direction in reciprocal space, in contrast to our findings for the CDW linewidths (see Supplementary Materials for more details). From this symmetry analysis, we can conclude that for both uniaxial and biaxial anisotropy it is in principle possible \cite{DeltaQ_note} to discriminate between a pure checkerboard and a pure stripe charge order, even in presence of a distribution of canted domains (see Tables\,S1-S3 for a complete classification). Ultimately, the inequivalence of the peak broadening $\Delta Q$ along different directions for all studied YBCO samples, combined with the macroscopic ${C}_{4}$ symmetry breaking, provide a clear cut evidence for the unidirectional (stripe) intrinsic nature of the charge order \cite{Charge_note}.

\begin{figure}[h!]
\centering
\includegraphics[width=0.8\linewidth]{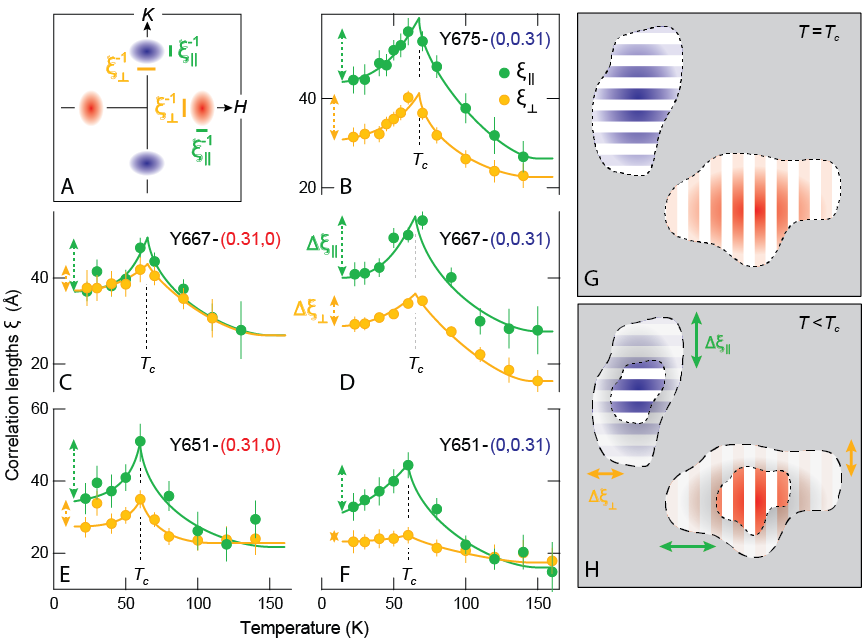}
\caption{\textbf{Interplay between stripe order and the superconducting state.} \textbf{(A)} Schematics of the CDW structure factor in Q-space, illustrating how longitudinal (${\xi}_{\parallel}$) and transverse (${\xi}_{\perp}$) correlation lengths are extracted from the data by taking the inverse of the peak HWHM $\Delta Q$. \textbf{(B-F)} Temperature dependence of ${\xi}_{\parallel}$ (green markers) and ${\xi}_{\perp}$ (orange markers) for the various samples and CDW peaks investigated (see legend). Continuous lines are guides-to-the-eye; double arrows highlight the drop in correlation lengths $\Delta {\xi}_{\parallel}$ and $\Delta {\xi}_{\perp}$ below ${T}_{\mathrm{c}}$. \textbf{(G,H)} Cartoons representing the evolution of the stripy nanodomains from above (G) to below (H) ${T}_{\mathrm{c}}$, illustrating how longitudinal correlations undergo a larger suppression in the presence of superconducting order.}
\label{Fig3}
\end{figure}

Having established the underlying stripe-like character of charge modulations in the CuO${}_{2}$ planes, we turn to the temperature dependence of the \textit{longitudinal} and \textit{transverse} correlation lengths, respectively parallel and perpendicular to the specific ordering wavevector. These can be extrapolated by inverting the momentum HWHM $\Delta Q$, as illustrated in Fig.\,\ref{Fig3}A. Longitudinal (transverse) correlations are then given by ${\xi}_{\parallel}\!=\! {\Delta Q}^{-1}_{\parallel}$ (${\xi}_{\perp}\!=\! {\Delta Q}^{-1}_{\perp}$), where ${\Delta Q}_{\parallel}$ (${\Delta Q}_{\perp}$) represents the momentum linewidth in the direction parallel (perpendicular) to the ordering wavevector. We subsequently study the temperature dependence of ${\xi}_{\parallel}$ and ${\xi}_{\perp}$ for both the ${\mathbf{Q}}_{a}$ and ${\mathbf{Q}}_{b}$ ordering wavevectors (Figs.\,\ref{Fig3}, B-F). First we observe the rise of correlation lengths below the CDW onset near 150\,K, followed by their suppression below the SC transition temperature ${T}_{\mathrm{c}}$, which confirms the competition between these two orders, in agreement with recent energy-integrated as well as energy-resolved RXS studies \cite{ghiringhelli2012,chang2012,achkar2012,Blanco2013,Blackburn2013}. However, the drop in the correlation lengths below ${T}_{\mathrm{c}}$ ($\Delta \xi$) is in all instances larger for the longidutinal correlations, or $\Delta {\xi}_{\parallel} > \Delta {\xi}_{\perp}$. In particular, the discrepancy between $\Delta {\xi}_{\parallel}$ and $\Delta {\xi}_{\perp}$, while small for Y675, is quite substantial for the more underdoped Y667 and Y651. This anisotropy provides additional evidence for the unidirectional nature of the charge ordering and thus the breaking of ${C}_{4}$ symmetry, since a bidirectional order would exhibit an isotropic drop in correlation length across ${T}_{\mathrm{c}}$. This anisotropy has an opposite doping trend from the crystal orthorhombicity whose associated anomalies across ${T}_{\mathrm{c}}$ increase with hole doping [as revealed e.g. by lattice espansivity measurements \cite{Meingast_2001}]. The inferred real-space representation of the evolution across ${T}_{\mathrm{c}}$ is schematically illustrated in Figs.\,\ref{Fig3}, G and H, where nanodomains are used to pictorially represent a charge-ordered state with finite correlation lengths. We conclude that the largest change occurs along the direction perpendicular to the stripes. This reflects the tendency of the SC order to gain strength, as temperature is lowered, primarily at the expense of longitudinal CDW correlations, suggesting that the mechanism responsible for the density fluctuations \textit{across} the periodically modulated stripes might be the main one competing with the Cooper pairing process. 

Our results may explain many common aspects between CDW physics in YBCO and the stripy cuprates from the La-based family, such as: thermoelectric transport \cite{Chroniere2009,laliberte2011}; strength of the order parameter \cite{Thampy_2013}; out-of-equilibrium response \cite{Torchinsky_2013,Hinton_2013}; energy-dependent RXS response \cite{achkar2012,achkar2013}. The nanoscopic nature of the stripe instability and the presence of both \textbf{a}- and \textbf{b}-oriented domains also clarify why this broken symmetry has been difficult to disentangle from a native bi-directional order \cite{ghiringhelli2012,chang2012,leboeuf_2014}, therefore requiring a tailored experimental scheme to resolve the 2D CDW structure factor $S ({Q}_{x},{Q}_{y})$. The pronounced directionality in the competition between superconductivity and stripe order reveals new key aspects of the underlying interplay between particle-particle and particle-hole pairing in high-temperature superconductors, and provides novel insights for an ultimate understanding of these materials.

\providecommand{\noopsort}[1]{}\providecommand{\singleletter}[1]{#1}%

\subsection*{Acknowledgments}

\normalsize{We acknowledge P. Abbamonte, E. Blackburn, L. Braicovich, J. Geck, G. Ghiringhelli, B. Keimer, S. Kivelson, M. Le Tacon, and S. Sachdev for insightful discussions. This work was supported by the Max Planck -- UBC Centre for Quantum Materials, the Killam, Alfred P. Sloan, Alexander von Humboldt, and NSERC's Steacie Memorial Fellowships (A.D.), the Canada Research Chairs Program (A.D., G.A.S.), NSERC, CFI, and CIFAR Quantum Materials. All of the experiments were performed at beamline REIXS of the Canadian Light Source, which is funded by the CFI, NSERC, NRC, CIHR, the Government of Saskatchewan, WD Canada, and the University of Saskatchewan. R.C. acknowledges the receipt of support from the CLS Graduate Student Travel Support Program. E.H.d.S.N acknowledges support from the CIFAR Global Academy.}



\clearpage

\clearpage

\begin{center}

\subsection*{Supplementary Materials\\Broken translational and rotational symmetry via charge stripe order in underdoped YBa$_{2}$Cu$_3$O$_{6+y}$} 

\large{R. Comin,$^{\ast,1,2}$ R. Sutarto,$^{3}$ E. H. da Silva Neto,$^{1,2,4,5}$ L. Chauviere,$^{1,2,5}$\\R. Liang,$^{1,2}$ W. N. Hardy,$^{1,2}$ D. A. Bonn,$^{1,2}$ F. He,$^{3}$ G. A. Sawatzky,$^{1,2}$\\and A. Damascelli$^{\ast,1,2}$}

\vspace{1cm}

\normalsize{$^{1}$Department of Physics {\rm {\&}} Astronomy, University of British Columbia,}\\
\normalsize{Vancouver, British Columbia V6T\,1Z1, Canada}\\
\normalsize{$^{2}$Quantum Matter Institute, University of British Columbia,}\\ \normalsize{Vancouver, British Columbia V6T\,1Z4, Canada}\\
\normalsize{$^{3}$Canadian Light Source, Saskatoon, Saskatchewan S7N\,2V3, Canada}\\
\normalsize{$^{4}$Quantum Materials Program, Canadian Institute for Advanced Research,}\\ \normalsize{Toronto, Ontario M5G 1Z8, Canada}\\
\normalsize{$^{5}$Max Planck Institute for Solid State Research,}\\
\normalsize{Heisenbergstrasse 1, D-70569 Stuttgart, Germany}\\
\vspace{1cm}
\normalsize{$^\ast$To whom correspondence should be addressed.}\\
\normalsize{E-mail: r.comin@utoronto.ca (R.C.); damascelli@physics.ubc.ca (A.D.)}

\end{center}

\makeatletter
\renewcommand{\fnum@table}{\tablename~S\thetable}
\setcounter{table}{0}
\makeatother
\makeatletter
\renewcommand{\fnum@figure}{\figurename~S\thefigure}
\setcounter{figure}{0}
\makeatother

\renewcommand{\theequation}{S\arabic{equation}} 

\clearpage

\subsection*{Materials and Methods}

\noindent {\bf Sample characterization.} This investigation was performed on three detwinned, oxygen-ordered underdoped YBa$_{2}$Cu$_3$O$_{6+y}$ single crystals: (i) $y\!=\!0.51$, $p\!\simeq\!0.10$, ${T}_{\mathrm{c}}\!=\!57$\,K, Ortho\,II oxygen-ordered, hereafter labeled Y651; (ii) $y\!=\!0.67$, $p\!\simeq\!0.12$, ${T}_{\mathrm{c}}\!=\!65$\,K, Ortho\,VIII oxygen-ordered, hereafter labeled Y667; (iii) $y\!=\!0.75$, $p\!\simeq\!0.13$, ${T}_{\mathrm{c}}\!=\!70$\,K, Ortho\,III oxygen-ordered, hereafter labeled Y675. The in-plane lattice constants, which in turn determine the degree of orthorhombicity, are: (i) $a\!\simeq\!3.8362 \AA$ and $b\!\simeq\!3.8740 \AA$ in Y651; (ii) $a\!\simeq\!3.8257 \AA$ and $b\!\simeq\!3.8814 \AA$ in Y667; and (iii) $a\!\simeq\!3.825 \AA$ and $b\!\simeq\!3.885 \AA$ in Y675. The superconducting critical temperature $T_{\mathrm{c}}$ was determined from magnetic susceptibility measurements. The in-plane lattice constants have been measured using XRD in Y651 and Y667, and are taken from Ref.\,44 for Y675. Details on the sample growth and the $T_{\mathrm{c}}$-to-doping correspondence can be found in Ref.\,45.
\\

\noindent {\bf Resonant soft X-ray scattering.} Resonant x-ray scattering (RXS) is a coherent probe of fluctuations in those physical quantities ($X$) that are directly or indirectly coupled to light, and is therefore proportional to the Fourier transform of the correlation function of $\delta X$, i.e.:
\begin{equation}
{I}_{RXS} (\mathbf{Q}) \propto \int d\mathbf{r} {e}^{-i \mathbf{Q} \cdot \mathbf{r}} \int d{\mathbf{r}}^{\prime} \left\langle {\delta X ({\mathbf{r}}^{\prime})\delta  X ({\mathbf{r}}^{\prime} + \mathbf{r})} \right\rangle = \left\langle { \delta X (- \mathbf{Q}) \delta X (\mathbf{Q})} \right\rangle, 
\end{equation}
where ${C}_{X} (\mathbf{r}) = \int d{\mathbf{r}}^{\prime} \left\langle {\delta X ({\mathbf{r}}^{\prime})\delta  X ({\mathbf{r}}^{\prime} + \mathbf{r})} \right\rangle $ represents the correlation function of $\delta X$.

Charge ordering is manifested through variations in different physical quantities, such as the ionic displacements ($\mathbf{X} = \delta \mathbf{R}$, in this case \textit{X} is a vector operator), the local valence ($X = p$), or the energy shifts of the resonant transition ($X = \Delta E$). The sensitivity of the experimental signal to these fluctuating quantities is material-dependent: in the cuprates it has been shown that the resonant scattering is predominantly controlled by variations in the energy shifts (i.e., a scalar field) [30,31], whereas in a different material, e.g. IrTe${}_{2}$, the RXS signal (at the Te-\textit{M} absorption edge) was shown to be primarily associated to a combination of valence modulations and atomic displacements [46]. In presence of valence modulations or energy shifts (i.e., scalar operators), the fluctuating quantity is proportional to modulations in the electronic density $\rho$, i.e. $\delta X \propto \delta \rho$, which leads to:
\begin{equation}
{I}_{RXS} (\mathbf{Q}) \propto \left\langle { \delta \rho (- \mathbf{Q}) \delta \rho (\mathbf{Q})} \right\rangle = \left\langle \delta {{\rho}^{*}} (\mathbf{Q}) \delta \rho (\mathbf{Q}) \right\rangle = S(\mathbf{Q}),
\label{eq:RXS_OP_relation}
\end{equation}
\noindent
where the second equality follows from the fact that the density operator is a real function, and $S (\mathbf{Q})$ is termed the \textit{static structure factor}.

\begin{figure}[h!]
\centering
\includegraphics[width=0.5\linewidth]{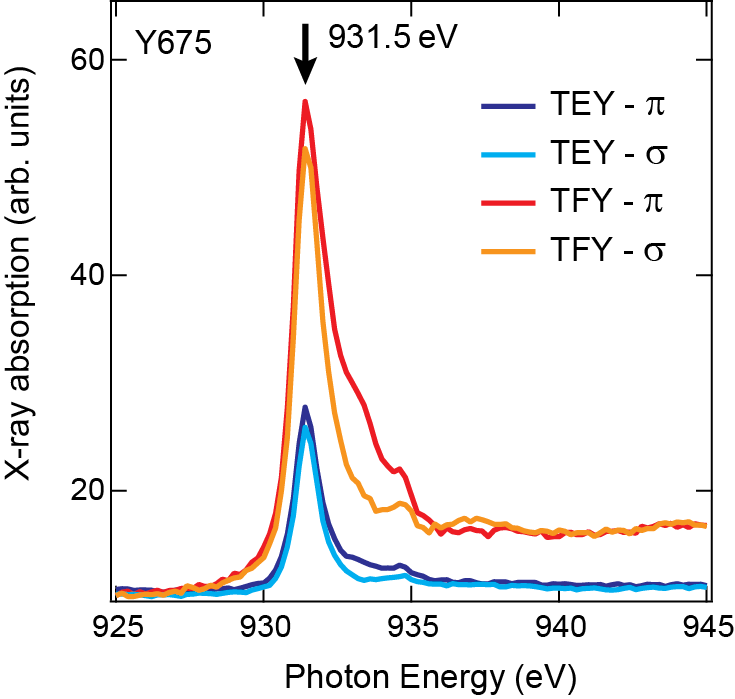}
\caption{\textbf{X-ray absorption at the Cu-$\mathbf{{L}_{3}}$ edge in Y675.} Shown are the x-ray absorption profiles measured in both total fluorescence yield (TFY) and total electron yield (TEY) mode, for incoming vertical ($\sigma$) and horizontal ($\pi$) polarization. The arrow marks the photon energy of the main transition involving planar Cu atoms (maximizing the CDW signal, see Refs.\,10,30), where all measurements were performed.}
\label{XAS}
\end{figure}

For this work, RXS measurements were performed on a 4-circle diffractometer in a ${10}^{-10}$\,mbar ultra-high-vacuum chamber, with a photon flux around $5 \cdot {10}^{12}$\,photons/s and $\frac{\Delta E}{E}\!\sim\! 2 \cdot {10}^{-4}$ energy resolution. In addition, fully polarized incoming light is used, with two available configurations: $\sigma$ (polarization vector perpendicular to the scattering plane) or $\pi$ (polarization vector in the scattering plane). In order to maximize the charge order signal, the series of measurements as a function of the azimuthal angle $\alpha$ were performed at the peak energy of the Cu-${L}_{3}$ edge ($h \nu \!=\! 931.5$\,eV, see Fig.\,S\ref{XAS}), at a detector angle ${\theta}_{det} \!=\! {170}^{\circ}$, and at a temperature near the superconducting transition ${T}_{\mathrm{c}}$. All RXS scans were measured with a multi-channel-plate detector with an angular resolution $\Delta \theta \sim {0.2}^{\circ}$, corresponding to an equivalent momentum resolution $\Delta Q \sim 0.0023 {\AA}^{-1}$ at $h \nu \!=\! 931.5$\,eV.

One should note that the spectrometer integrates the energy of scattered photons; however, since inelastic process are \textit{incoherent} in nature, the only portion of the spectrum which carries the information on the momentum structure of the electronic density is the elastic (zero-loss) line. Since the inelastic contributions, albeit strong at or near resonance, do not have a sharp structure in momentum space, the momentum structure of the CDW peaks as observed with an energy-integrated spectrometer is still predominantly determined by the elastic processes, and therefore it is representative of the static electronic modulations in Fourier space. This is also confirmed by the close agreement between the CDW momentum linewidth previously measured in the energy-resolved (RIXS) and energy-integrated (RXS) mode (see Fig.\,4 in Ref.\,10).

\subsection*{Supplementary Text}

\begin{figure}[h!]
\centering
\includegraphics[width=1\linewidth]{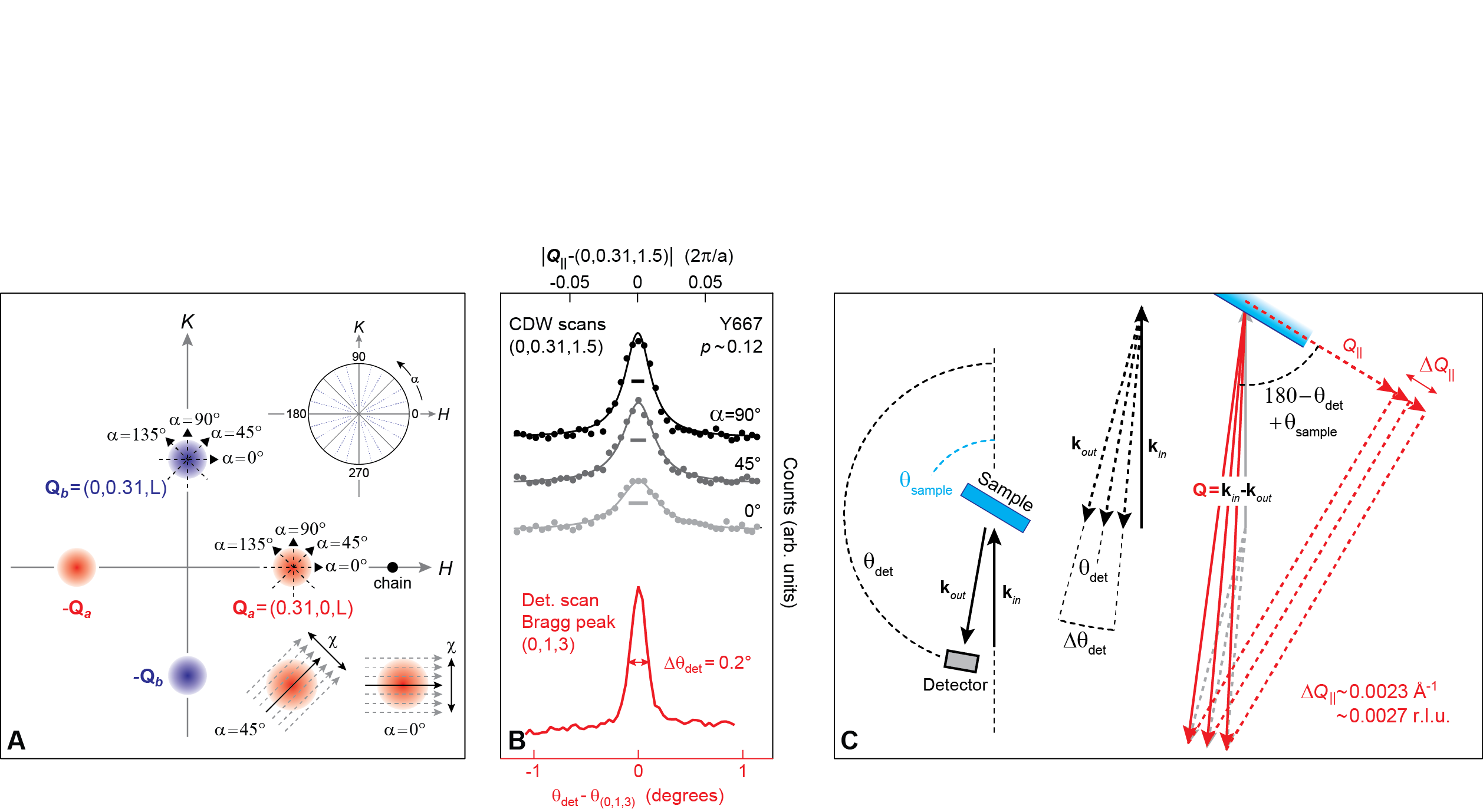}
\caption{\textbf{Azimuthal geometry and angular resolution.} \textbf{(A)} Schematics of momentum-resolved electronic density in YBCO systems, highlighting: (i) the chain-ordering feature (black dot), which in detwinned samples is only located along the reciprocal $H$ axis; (ii) plane-ordering CDW peaks at $\pm {\mathbf{Q}}_{a}\!=\!(\pm 0.31,0)$ along $H$ (red diffuse circles); (iii) plane-ordering CDW peaks at $\pm {\mathbf{Q}}_{b}\!=\!(0, \pm 0.31)$ along $K$ (blue diffuse circles). The zero of the azimuthal angle ($\alpha$) scale defines a generic direction in reciprocal space which is parallel to the $H$ axis (top-right inset). Also shown are the Q-cuts for various values of $\alpha$ around the two CDW peaks ${\mathbf{Q}}_{a}$ and ${\mathbf{Q}}_{b}$. \textbf{(B)} The top part of the graph reproduces Fig.\,1a from the main text, showing the RXS scans (markers) across the CDW peak ${\mathbf{Q}}_{b}\!=\!(0,0.31)$ as a function of $K$, with Lorentzian fits overlaid. The bottom part (red curve) shows a scan of the photodetector across the Bragg peak ${Q}_{\mathrm{Bragg}}\!=\!(0,1,3)$ as a function of detector angle ${\theta}_{\mathrm{det}}$. The angular resolution, as obtained by fitting the red curve with a Gaussian lineshape, is equal to $\Delta {\theta}_{\mathrm{det}}\!=\!{0.2}^{\circ}$. \textbf{(C)} RXS probing geometry highlighting how the surface-projected momentum resolution $\Delta {Q}_{\parallel}$ is derived starting from the angular resolution $\Delta {\theta}_{\mathrm{det}}$.}
\label{Azimuth_geometry}
\end{figure}

\noindent {\bf Azimuthal scans and experimental geometry.} The azimuthal angle $\alpha$ is defined as the angle between the direction of the RXS scan in momentum space and the crystallographic \textbf{a} axis, or equivalently its reciprocal axis $H$. This configuration is illustrated in Fig.\,S\ref{Azimuth_geometry}A, which also highlights the cuts in momentum space (for different $\alpha$ values), at the charge-density-wave (CDW) ordering wavevectors studied here, namely ${\mathbf{Q}}_{a}\!=\!(0.31,0,1.5)$ and ${\mathbf{Q}}_{b}\!=\!(0,0.31,1.5)$. The value of $L\!=\!1.5$ is chosen to maximize the CDW intensity [10]. However, as studied in detail in recent work [11,24,26], the dependence of the CDW ordering peak on the out-of-plane wavevector component \textit{L} is very weak, a signature of the two-dimensional nature of the charge order. For this reason, only the planar momentum structure of the CDW order parameter is studied here, and we will henceforth refer exclusively to the in-plane projection of the CDW peaks, and use the notation ${\mathbf{Q}}_{a}\!=\!(0.31,0)$ and ${\mathbf{Q}}_{b}\!=\!(0,0.31)$.

In order to extract a proper profiling of the CDW peak, it is important that each Q-scan cuts through the CDW peak maximum. To ensure that this condition is met for each azimuthal angle $\alpha$, angular optimization perpendicular to the Q-scan direction was performed by scanning through the peak as a function of the transverse angle $\chi$ (see lower-left inset of Fig.\,S\ref{Azimuth_geometry}A). By finding the $\chi$ angle maximizing the CDW signal, we have been able to cut precisely through the peak maximum for every azimuthal position. 

In a RXS measurement, the angle-to-momentum conversion is encoded in the equations:
\begin{align}
Q ({\theta}_{\mathrm{det}}, h \nu) &= \frac{4 \pi}{c \nu} \sin \left( \frac{{\theta}_{\mathrm{det}}}{2} \right),\nonumber\\ {Q}_{\parallel} ({\theta}_{\mathrm{det}}, {\theta}_{\mathrm{sample}}, h \nu) &= \frac{4 \pi}{c \nu} \sin \left( \frac{{\theta}_{\mathrm{det}}}{2} \right) \cos \left( \pi - {\theta}_{\mathrm{det}} + {\theta}_{\mathrm{sample}} \right),
\label{eq:RXS_formulae}
\end{align}
where $Q$ is the wavevector magnitude, ${Q}_{\parallel}$ is its projection on the sample surface ($\mathbf{a}-\mathbf{b}$ plane), while ${\theta}_{\mathrm{det}}$ is the detector angle, ${\theta}_{\mathrm{sample}}$ is the sample angular position (see schematics of the probing geometry in Fig.\,\ref{Azimuth_geometry}C), while $\nu$ is the photon frequency. For the charge ordering peak, the angular values are the following: ${\theta}_{\mathrm{det}} \!=\! {170}^{\circ}$, ${\theta}_{\mathrm{sample}} \!=\! {32.5}^{\circ}$.

The resolution of the RXS diffractometer is evaluated by measuring the Bragg reflection $(0,1,3)$ at high photon energy ($h \nu \!=\! 2230$\,eV). By scanning the detector arm across the Bragg reflection -- whose intrinsic broadening is much smaller than the instrumental resolution and therefore represents a $\delta$-like signal -- we estimate the angular resolution to be $\Delta {\theta}_{\mathrm{\det}} \sim {0.2}^{\circ}$ (see lower part of Fig.\,S\ref{Azimuth_geometry}B).  In order to estimate the momentum resolution, we have converted the angular resolution scan (see again bottom part of Fig.\,S\ref{Azimuth_geometry}B) into the surface projected wavevector and then evaluated $\Delta {Q}_{\parallel}$ using Eqs.\,\ref{eq:RXS_formulae} as follows:
\begin{align}
\Delta {Q}_{\parallel} &= {Q}_{\parallel} \left( {\theta}_{\mathrm{det}} + \Delta {\theta}_{\mathrm{det}}/2, {\theta}_{\mathrm{sample}}, h \nu \right) - {Q}_{\parallel} \left( {\theta}_{\mathrm{det}} - \Delta {\theta}_{\mathrm{det}}/2, {\theta}_{\mathrm{sample}}, h \nu \right)\nonumber\\
&= {Q}_{\parallel} \left( {170}^{\circ} + {0.1}^{\circ}, {32.5}^{\circ}, 930.5 eV \right) - {Q}_{\parallel} \left( {170}^{\circ} - {0.1}^{\circ}, {32.5}^{\circ}, 930.5 eV \right)\nonumber\\
&= \frac{4 \pi}{c \nu} \left[ \sin \left( {85.05}^{\circ} \right) \cos \left( {180-170.1+32.5}^{\circ} \right) - \sin \left( {84.95}^{\circ} \right) \cos \left( {180-169.9+32.5}^{\circ} \right) \right]\nonumber\\
&= 0.694668 - 0.692344 \sim 0.0023 {\AA}^{-1}.
\end{align}
The value obtained amounts to approximately the 5\% of the average full-width-at-half-maximum for the charge order peak (which is of the order of $0.04-0.05 {\AA}^{-1}$). Since the two contributions add in quadrature, if we use the term ${\Delta Q}_{\mathrm{native}}$ to denote the intrinsic contribution as opposed to the one from resolution, ${\Delta Q}_{\mathrm{res}}$, then we have for the measured linewidths that ${\Delta Q}^{2}_{\mathrm{exp}} \!=\! {\Delta Q}^{2}_{\mathrm{native}} + {\Delta Q}^{2}_{\mathrm{res}}$. Since ${\Delta Q}_{\mathrm{res}} \sim 0.05 \cdot {\Delta Q}_{\mathrm{native}}$, we obtain that ${\Delta Q}_{\mathrm{native}} \!=\! 0.9987 \cdot {\Delta Q}_{\mathrm{exp}}$, which shows that approximately the 99.8\% of the observed peak broadening (see again Fig.1 of the main text) represents intrinsic sample properties. 
\\

\noindent {\bf Data analysis methodology.} RXS scans have been analyzed by means of a least-square nonlinear regression analysis, with fitting function defined as the sum of a Lorentzian peak and a cubic background polynomial function \textit{B}:
\begin{align}
{I}_{\mathrm{RXS}} &= \frac{A}{{\left(\dfrac{Q-{Q}^{*}}{\Delta Q}\right)}^{2} + 1} + B (Q)\nonumber\\
B (Q) &= {a}_{0} + {a}_{1} \: Q + {a}_{2} \: {Q}^{2} + {a}_{3} \: {Q}^{3},
\label{eq:fit_function}
\end{align}
where ${Q}^{*}$ is the peak position, $A$ is the amplitude and $\Delta Q$ is the half-width-at-half-maximum.

\begin{figure}[t!]
\centering
\includegraphics[width=0.95\linewidth]{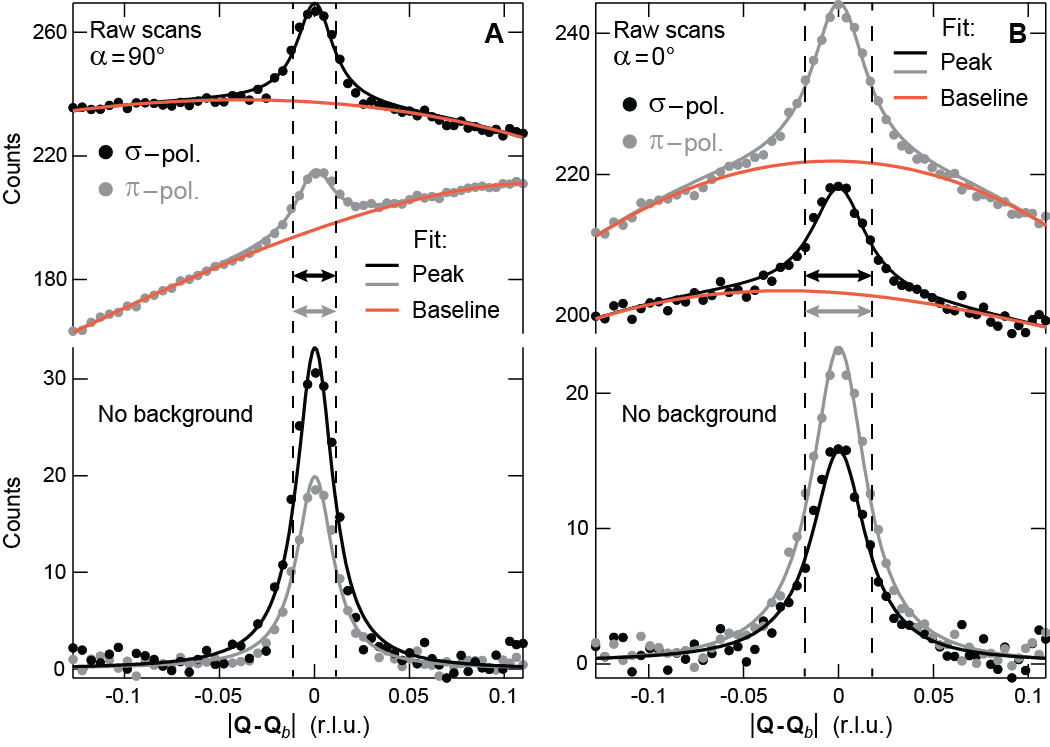}
\caption{\textbf{RXS data analysis.} \textbf{(A)} Top part: RXS scans acquired on a Y667 sample around ${\mathbf{Q}}_{b}\!=\!(0,0.31,1.5)$ and at $\alpha \!=\! {90}^{\circ}$ using vertical ($\sigma$) light polarization (black markers) and horizontal ($\pi$) light polarization (grey markers). Fitted profiles based on Eq.\,\ref{eq:fit_function} are shown as continuous lines, while the cubic baseline is shown in orange; double arrows denote the full-width-at-half-maximum, which is constrained to be equal for both scans. Bottom part: Background-subtracted scans, showing only the resonant contribution from the density modulations. \textbf{(B)} Same as \textbf{(A)}, but acquired at an azimuthal position $\alpha \!=\! {0}^{\circ}$.}
\label{Fitting_example}
\end{figure}

All momentum scans were measured around the ordering wavevector using two light polarizations, vertical ($\sigma$) and horizontal ($\pi$). A preliminary analysis has shown that the linewidths do not depend on the light polarization, consistently with the fact that the linewidth is an \textit{intrinsic} property of the system (related to correlation lengths, as will be discussed later) and therefore does not vary upon changing an \textit{external} parameter, such as light polarization. In order to provide a more precise quantitative estimate of the linewidth $\Delta Q$ with a reduced statistical error, this parameter has been subsequently constrained to be the same during least-squares fitting of the $\sigma$- and $\pi$-polarized scans. An example of the fitting analysis is provided in Fig.\,S\ref{Fitting_example}.
\\

\noindent {\bf Stripe order with anisotropic correlations.} As anticipated in the Materials and Methods section, in our case the RXS observable is proportional to the static structure factor $S (\mathbf{Q})$:
\begin{equation}
{I}_{\mathrm{RXS}} (\mathbf{Q}) \propto S (\mathbf{Q}) = \left\langle {\left\vert \delta \rho (\mathbf{Q}) \right\vert}^{2} \right\rangle.
\end{equation}
\noindent
Here we focus on how a rotational symmetry breaking in real space -- a tell-tale of a stripe-like charge-ordered state -- can be equivalently detected by looking at the structure of density modulations in reciprocal space. Whenever ${C}_{4}$ symmetry is preserved, the electronic density fluctuations $\delta \rho$ must be invariant under a ${90}^{\circ}$ rotation in real space (defined through the transformations $x \rightarrow y$ and $y \rightarrow -x$), or $\delta \rho (x,y) \!=\! \delta \rho (y,-x)$. This can be readily translated in momentum space by taking the Fourier transform of $\delta \rho (x,y)$, which yields:
\begin{align}
\delta \rho ({Q}_{x},{Q}_{y}) &= \int \mathrm{d} x \: \mathrm{d} y \: {e}^{-i \left( {Q}_{x} x + {Q}_{y} y \right)} \delta \rho (x,y) \nonumber\\ &= \int \mathrm{d} x \: \mathrm{d} y \: {e}^{-i \left( {Q}_{x} x + {Q}_{y} y \right)} \delta \rho (y,-x) \nonumber\\ &= \int \mathrm{d} {x}^{\prime} \: \mathrm{d} {y}^{\prime} \: {e}^{-i \left( -{Q}_{x} {y}^{\prime} + {Q}_{y} {x}^{\prime} \right)} \delta \rho ({x}^{\prime},{y}^{\prime}) \nonumber\\ &= \delta \rho ({Q}_{y},-{Q}_{x}),
\label{eq:S_Q_rotation}
\end{align}
which proves that the structure factor $S (\mathbf{Q})$ has to likewise be invariant under a ${90}^{\circ}$ rotation in momentum space, or $S ({Q}_{x},{Q}_{y}) \!=\! S ({Q}_{y},-{Q}_{x})$. We then apply this procedure to the full 2D momentum structure of the CDW structure factor as measured using RXS (see Fig.\,1 in the main text). 

\begin{figure}[t!]
\centering
\includegraphics[width=0.9\linewidth]{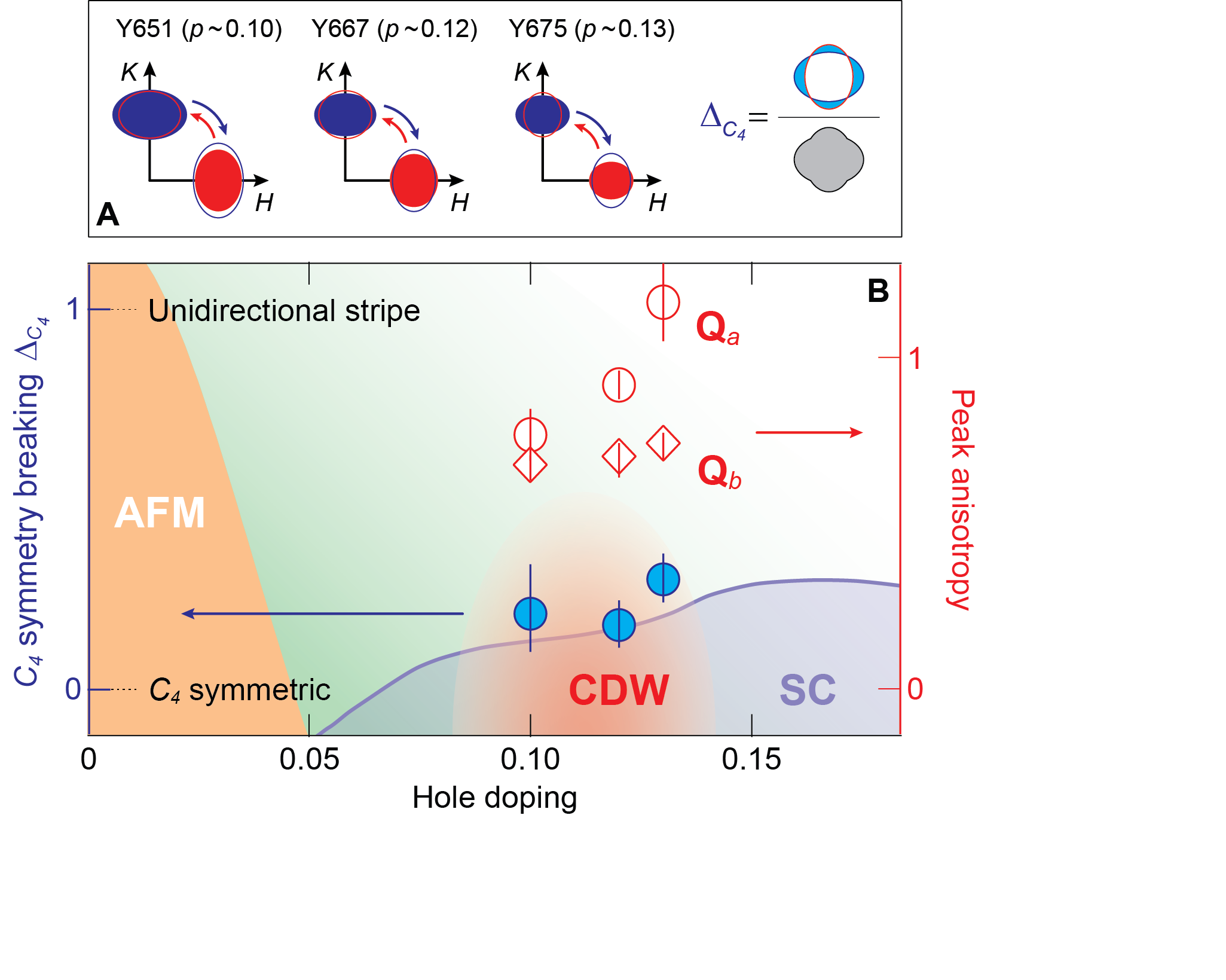}
\caption{\textbf{Momentum-space evidence of rotational symmetry breaking.} \textbf{(A)} Schematic representation of the comparison between the CDW structure in momentum (full ellipses) and its rotated version (hollow ellipses), clearly showing a violation of Eq.\,\ref{eq:S_Q_rotation}. Right panel: graphical definition of ${\Delta}_{{C}_{4}}$ as the ratio between the light blue and grey areas as obtained from the original and rotated CDW structures shown in \textbf{(a)}, taking as example the specific case of Y675. $ {\Delta}_{{C}_{4}}$ is formally defined as ${\Delta}_{{C}_{4}} \!=\! \left[ S({\mathbf{Q}}_{a}) \cup \tilde{S} ({\mathbf{Q}}_{b}) - S({\mathbf{Q}}_{a}) \cap \tilde{S} ({\mathbf{Q}}_{b}) \right] / \left[ S({\mathbf{Q}}_{a}) \cup \tilde{S} ({\mathbf{Q}}_{b}) \right] $, or equivalently as the area of the union of the two CDW peaks (one of which rotated) $S({\mathbf{Q}}_{a}) \cup \tilde{S} ({\mathbf{Q}}_{b})$ minus their overlap area $S({\mathbf{Q}}_{a}) \cap \tilde{S} ({\mathbf{Q}}_{b})$, divided by their union. \textbf{(B)} Doping dependence of ${\Delta}_{{C}_{4}}$ (blue markers) and of the peak anisotropy (from the ratio between longitudinal and transverse peak width ${\Delta Q}_{\parallel} / {\Delta Q}_{\perp}$) for both CDW peaks ${\mathbf{Q}}_{a}$ (dark red) and ${\mathbf{Q}}_{b}$ (red).}
\label{Symmetry_OP_rot}
\end{figure}

The resulting peak structure is readily shown to violate ${C}_{4}$ symmetry, as illustrated in the three panels of Fig.\,S\ref{Symmetry_OP_rot}A (reproduced from Fig.\,1 of the main text). The extent of the rotational symmetry breaking is here quantified via a ${C}_{4}$ symmetry breaking parameter ${\Delta}_{{C}_{4}}$, defined as ${\Delta}_{{C}_{4}} \!=\! \left[ S({\mathbf{Q}}_{a}) \cup \tilde{S} ({\mathbf{Q}}_{b}) - S({\mathbf{Q}}_{a}) \cap \tilde{S} ({\mathbf{Q}}_{b}) \right] / \left[ S({\mathbf{Q}}_{a}) \cup \tilde{S} ({\mathbf{Q}}_{b}) \right] $, or equivalently as the area of the union of the two CDW peaks (one of which rotated) $S({\mathbf{Q}}_{a}) \cup \tilde{S} ({\mathbf{Q}}_{b})$ minus their overlap area $S({\mathbf{Q}}_{a}) \cap \tilde{S} ({\mathbf{Q}}_{b})$, divided by their union. The doping dependence of ${\Delta}_{{C}_{4}}$ is then shown in Fig.\,S\ref{Symmetry_OP_rot}B. On the same plot, we report the values of the peak anisotropy ${\gamma}_{a,b}$ for both the ${\mathbf{Q}}_{a}$ and ${\mathbf{Q}}_{b}$ CDW peaks, as derived by taking the ratio between the peak width along the parallel and perpendicular direction with respect to the ordering wavevector, or ${\gamma} \!=\! {\Delta Q}_{\parallel} / {\Delta Q}_{\perp}$. A value of $\gamma \!=\! 1$ yields isotropic peaks, while the larger the deviation from 1, the more anisotropic the peak is (in one or the other direction). Interestingly, for Y651 and Y667, both ${\gamma}_{a}$ and ${\gamma}_{b}$ are smaller than 1, corresponding to CDW peak structures elongated differently between the two directions in reciprocal space. On the other hand, in Y675 we find that ${\gamma}_{a} \!>\! 1$, ${\gamma}_{b} \!<\! 1$, corresponding to having ${\mathbf{Q}}_{a}$ and ${\mathbf{Q}}_{b}$ elongated along the same direction (see again Fig.\,S\ref{Symmetry_OP_rot}A). Herefater we will use the denomination \textit{biaxial anisotropy} for the case of Y651 and Y667, and \textit{uniaxial anisotropy} for the case of Y675 (see more below on uni vs. biaxial).

More detailed doping-, azimuthal angle-, and temperature-dependent results on peak widths and correlation lengths are presented in the next two sections.
\\

\begin{figure}[t!]
\centering
\includegraphics[width=1\linewidth]{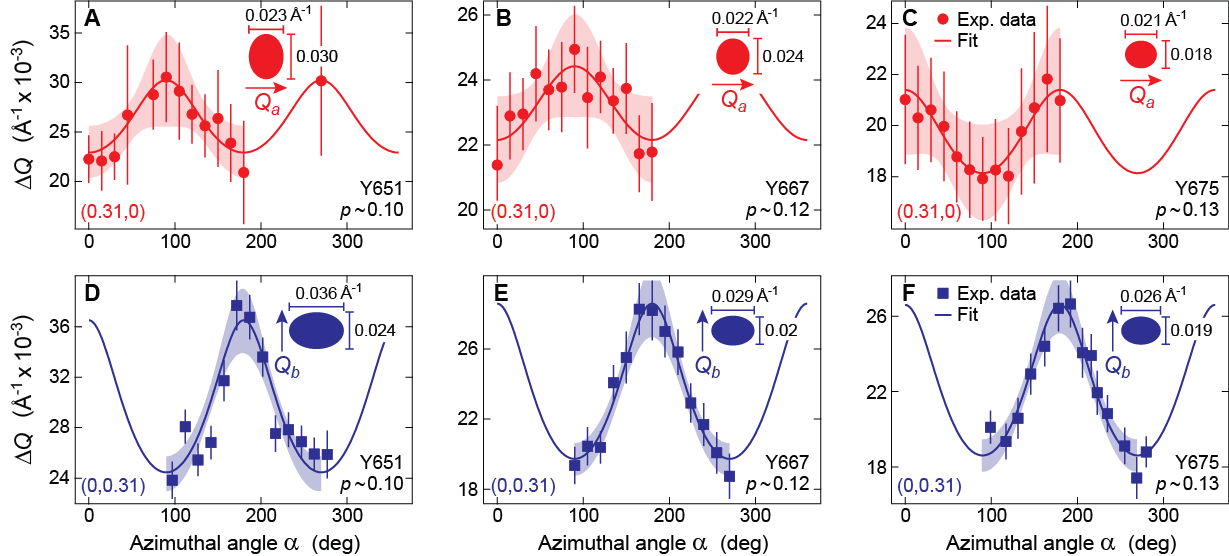}
\caption{\textbf{CDW peak linewidth vs. azimuthal angle.} \textbf{(A-C)} Plots of the half-width-at-half-maximum $\Delta Q$ of the CDW peak ${Q}_{a}\!=\!(0.31,0,1.5)$ as a function of azimuthal angle $\alpha$, for Y651, Y667, and Y675, respectively. \textbf{(D-F)} Same as \textbf{(A-C)}, but for the CDW peak ${Q}_{b}\!=\!(0,0.31,1.5)$. Experimental points (colored markers) are expressed in ${\AA}^{-1} \times {10}^{-3}$. The uncertainties (error bars) on the data points have been estimated from the fitting analysis of RXS scans using standard procedures [48]. Continuous lines are fits to the $\Delta Q (\alpha)$ data points using Eq.\,\ref{eq:Gamma_alpha}, while shaded areas represent the 95\% confidence bands, i.e. the range within which the model fit covers a 95\% probability of representing the true model, in the presence of the reported uncertainties on $\Delta Q$ (indicated by error bars). The inset diagrams report the values of $\Delta Q$ along the reciprocal axes $H$ and $K$, and illustrate the resulting peak shape in momentum space.}
\label{Qwidths_all}
\end{figure}

\noindent {\bf Azimuthal-dependent experimental data and parameters.} The two-dimensional structure factor near the ordering wavevectors $\mathbf{Q}_{a}$ and $\mathbf{Q}_{b}$ is well-approximated by a two-dimensional Lorentzian function defined as follows:
\begin{equation}
{I}^{2D}_{\mathrm{Lor}} \left( {Q}_{x}, {Q}_{y} \right) = \frac{A}{{\left(\dfrac{{Q}_{x}-{Q}^{*}_{x}}{{\Delta Q}_{x}}\right)}^{2} + {\left(\dfrac{{Q}_{y}-{Q}^{*}_{y}}{{\Delta Q}_{y}}\right)}^{2} + 1} = \frac{A}{{\left(\dfrac{{Q}_{x}^{\prime}}{{\Delta Q}_{x}}\right)}^{2} + {\left(\dfrac{{Q}_{y}^{\prime}}{{\Delta Q}_{y}}\right)}^{2} + 1},
\label{eq:I_tot}
\end{equation}
where $\mathbf{Q}^{*} \!=\! ({Q}^{*}_{x},{Q}^{*}_{y})$ is the CDW wavevector, ${\Delta Q}_{x,y}$ are the half-widths-at-half-maximum (equal to the inverse correlation lengths) along \textbf{x} and \textbf{y}, respectively, while in the last equality the reduced coordinate $\mathbf{Q}^{\prime} = \mathbf{Q} - \mathbf{Q}^{*}$ was used. By parametrizing $\mathbf{Q}^{\prime}$ in polar coordinates as $( {Q}_{x}^{\prime}, {Q}_{y}^{\prime}) = {Q}^{\prime} ( \cos \alpha, \sin \alpha ) $, we can subsequently study the directional dependence of the RXS intensity from Eq.\,\ref{eq:I_tot}, where the direction in reciprocal space is controlled by the angle $\alpha$:
\begin{equation}
{I}^{2D}_{\mathrm{Lor}} \left( Q, \alpha \right) = \frac{A}{ {Q}^{\prime 2} \left( \dfrac{{{\cos}^{2} \alpha}}{{\Delta Q}_{x}^{2}} + \dfrac{{{\sin}^{2} \alpha}}{{\Delta Q}_{y}^{2}} \right) + 1} = \frac{A}{ \dfrac{{Q}^{\prime 2}}{{\Delta Q}_{\alpha}^{2}} + 1}.
\label{eq:I_alpha}
\end{equation}
Eq.\,\ref{eq:I_alpha} is used to fit the azimuthal series of RXS scans, which allows extracting the experimental $\alpha$-dependent linewidths ${\Delta Q}_{\alpha}$ (error bars are computed from a least-squares fitting analysis), shown in Fig.\,S\ref{Qwidths_all}. The same equation also allows reformulating ${\Delta Q}_{\alpha}$ as a function of the two independent parameters ${\Delta Q}_{x}$ and ${\Delta Q}_{y}$, which correspond to the linewidth at $\alpha\!=\! {0}^{\circ}$ and ${90}^{\circ}$ degrees, respectively:
\begin{equation}
{\Delta Q}_{\alpha} = {\left( \dfrac{{{\cos}^{2} \alpha}}{{\Delta Q}_{x}^{2}} + \dfrac{{{\sin}^{2} \alpha}}{{\Delta Q}_{y}^{2}} \right)}^{-1}.
\label{eq:Gamma_alpha}
\end{equation}
This last formula is used to fit the $\alpha$-dependent linewidths displayed in Fig.\,S\ref{Qwidths_all}; the corresponding fit profiles are overlaid as continuous lines. The values of ${\Delta Q}_{x}$ (${\Delta Q}_{y}$) are reported beside the horizontal (vertical) bars in the top-right insets, which also define the peak shape (full ellipsoid) and corresponding ${Q}_{x}/{Q}_{y}$ anisotropy.

\begin{figure}[t!]
\centering
\includegraphics[width=1\linewidth]{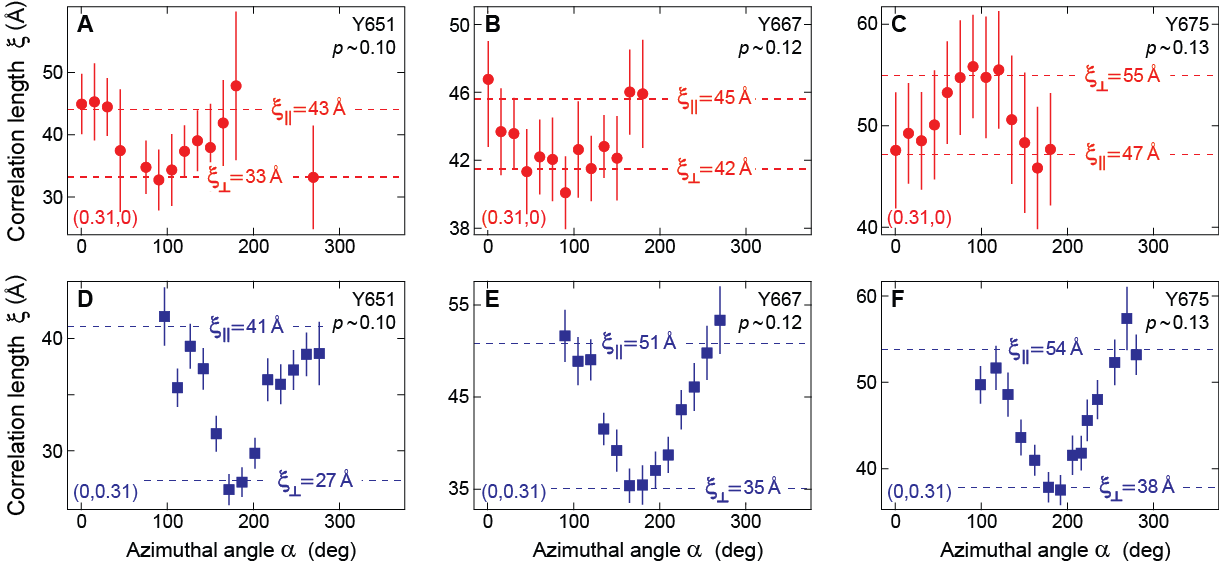}
\caption{\textbf{CDW correlation length vs. azimuthal angle.} \textbf{(A-C)} Plots of the CDW correlation length $\xi$ for ${Q}_{a}\!=\!(0.31,0,1.5)$ as a function of azimuthal angle $\alpha$, for Y651, Y667, and Y675, respectively. \textbf{(D-F)} Same as \textbf{(A-C)}, but for ${Q}_{b}\!=\!(0,0.31,1.5)$. Reported in each panel are the values of longitudinal (${\xi}_{\parallel}$) and transverse (${\xi}_{\perp}$) correlations. The uncertainties (error bars) on the data points have been derived from the corresponding uncertainties in the values of $\Delta Q$ (see Fig.\,\ref{Qwidths_all}) using error propagation, i.e. $\dfrac{{\delta}_{\Delta Q}}{\Delta Q} \!=\! \dfrac{{\delta}_{\xi}}{\xi}$.}
\label{Corrlens_all}
\end{figure}

From the values of ${\Delta Q}_{\alpha}$ we can derive the $\alpha$-dependent correlation lengths upon simple inversion: ${\xi}_{\alpha} \!=\! {\Delta Q}_{\alpha}^{-1}$. The corresponding data points are plotted in Fig.\,S\ref{Corrlens_all}. Here we also introduce longitudinal (${\xi}_{\parallel}$) and transverse (${\xi}_{\perp}$) correlations, representing the correlation lengths along the two axes that are respectively parallel and perpendicular to the specific wavevector. 
\\

\noindent {\bf Analysis and categorization of possible domain structures.} A more detailed study of the native anisotropy of the peak shape and consequently of the correlation lengths, in relation to possible CDW domain structures, is also important in revealing the stripy nature of charge ordering in YBCO. Such analysis relies on a few preliminary observations:
\begin{itemize}
\item The x-ray beam spot size ($d \!\sim\! 500 \mu m$) is much larger than the average correlation length ($\xi \!\sim\! 40-60 \AA$), hence variations in the latter with the azimuthal angle $\alpha$ cannot be attributed to the (minimal) motion of the beam spot on the probed region of the sample.
\item The shortest and longest correlations always occur at either $\alpha=0$ (corresponding to a direction parallel to the \textbf{a} axis) or $\alpha=90$ (parallel to the \textbf{b} axis) and therefore rule out the influence of extrinsic effects related to the probing geometry in determining the $\alpha$-dependence of the correlation lengths -- which would not preferentially select a high-symmetry direction, and would rather cause the signal to be minimized or maximized at a random angular position.
\item The different variation of $\xi$ with azimuthal angle between Y675 and the other two samples (see again Fig.\,S\ref{Corrlens_all}) also excludes any systematic effect to the probing geometry.
\end{itemize}

\begin{figure}[t!]
\centering
\includegraphics[width=1\linewidth]{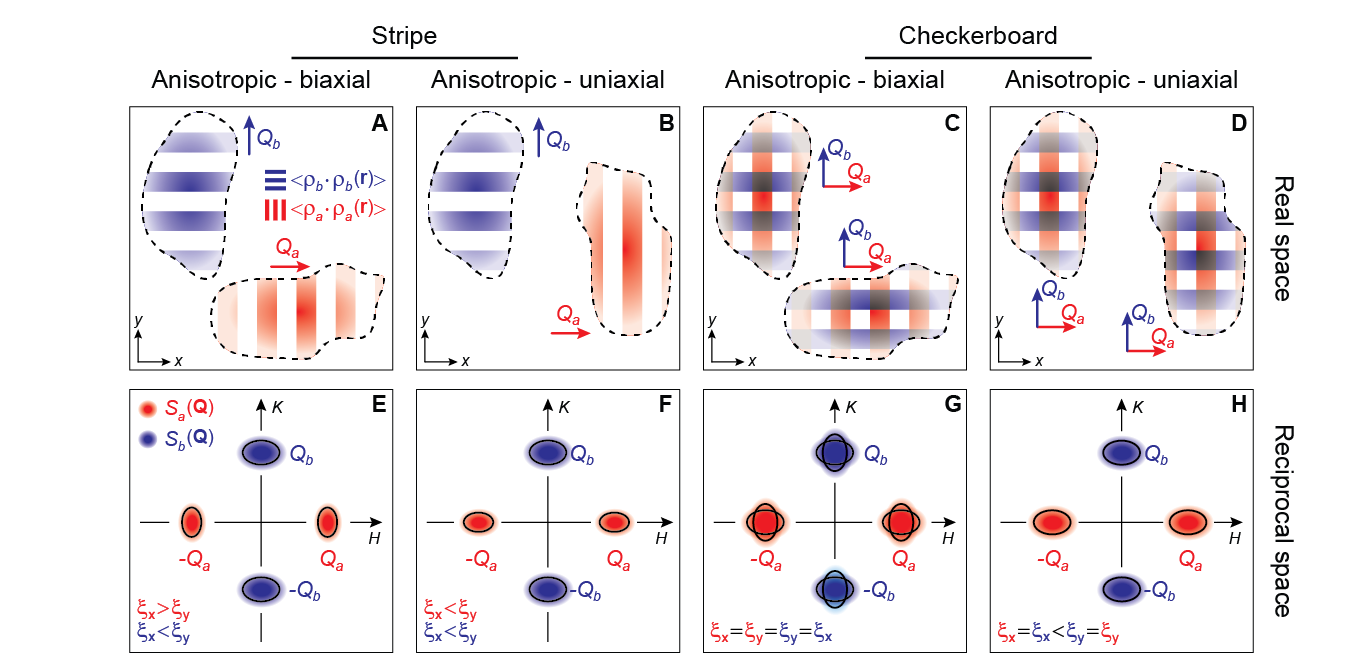}
\caption{\textbf{Stripe and checkerboard charge order parameters in presence of anisotropy.} \textbf{(A,B)} Stripe nanodomains in presence of biaxial \textbf{(A)} and uniaxial \textbf{(B)} anisotropic correlation lengths. \textbf{(C,D)} Checkerboard nanodomains in presence of biaxial \textbf{(C)} and uniaxial \textbf{(D)} anisotropic correlation lengths. \textbf{(E-H)} Corresponding structure factors in reciprocal (\textbf{Q}) space. In the case of uniaxial anisotropy \textbf{(F,H)}, a stripe and checkerboard order can be distinguished in Q-space provided that correlations along \textbf{x} or \textbf{y} are not equal between the two order parameter components, i.e. ${\xi}^{a}_{x} \neq {\xi}^{b}_{x}$, or ${\xi}^{a}_{y} \neq {\xi}^{b}_{y}$. In the case of biaxial anisotropy \textbf{(E,G)}, stripe order possesses an \textit{intra-domain} wavevector which can be locked to the correlation length, whereas this is not possible for the checkerboard case; in such case, stripe and checkerboard charge order can be distinguished even more clearly.}
\label{Stripe_CB_OP_all}
\end{figure}

Based on the results shown in Fig.\,S\ref{Corrlens_all}, anisotropic correlations are confirmed to be present in all YBCO samples examined. In general, we note how correlation lengths increase with increasing doping, whereas the type of anisotropy evolves from being biaxial for $p \!<\! 0.12$ to being uniaxial for $p \!>\! 0.12$. In order to establish a link to the domain meso-structure and show that the experimentally-found peak profiles are incompatible with microscopic bidirectional (checkerboard) order, we define the correlation lengths along the \textit{x} and \textit{y} (equivalently \textit{H} and \textit{K}) directions as ${\xi}^{a,b}_{x}$ and ${\xi}^{a,b}_{y}$, respectively (the superscript refers to the peak at ${\mathbf{Q}}_{a}$ or ${\mathbf{Q}}_{b}$). The latter quantities are related to the longitudinal and transverse correlations via the relations:
\begin{align}
{\xi}^{a}_{\parallel} &= {\xi}^{a}_{x} \qquad {\xi}^{a}_{\perp} = {\xi}^{a}_{y}\nonumber \\
{\xi}^{b}_{\parallel} &= {\xi}^{b}_{y} \qquad {\xi}^{b}_{\perp} = {\xi}^{b}_{x}
\end{align}
Therefore, for domains with wavevector ${\mathbf{Q}}_{a}\!=\!(0.31,0)$ [${\mathbf{Q}}_{b}\!=\!(0,0.31)$], ${\xi}_{\parallel}$ and ${\xi}_{\perp}$ are evaluated at $\alpha \!=\! {0}^{\circ}$ [${90}^{\circ}$] and $\alpha \!=\! {90}^{\circ}$ [${180}^{\circ}$] respectively. For all the investigated charge ordering peaks, we find that ${\xi}_{x} \neq {\xi}_{y}$. Whenever ${\xi}_{x} > {\xi}_{y}$ (${\xi}_{y} > {\xi}_{x}$), we will talk of x-oriented (y-oriented) domains, and we will similarly say that the anisotropy is along x (y). Since we are dealing with \textit{local}, not \textit{global} correlations, two situations may arise: (i) a \textit{biaxial anisotropy}, where both x- and y-oriented domains are present; (ii) a \textit{uniaxial anisotropy}, where only x-oriented (or y-oriented) are found. Case (ii) would effectively correspond to a \textit{global}, macroscopic anisotropy. These two possibilities are pictorially illustrated in Figs.\,S\ref{Stripe_CB_OP_all}A-D. In particular, the uniaxial anisotropy shown in Figs.\,S\ref{Stripe_CB_OP_all}B,D assumes the presence of y-oriented domains. Note that these domains need not necessarily lie in the very same layer, but they need to be present at the same time within the bulk of the material (i.e., they can be present in alternating layer, for instance). The momentum-space representation of the order parameter -- and therefore of the electronic density fluctuations is shown in the corresponding panels in Fig.\,S\ref{Stripe_CB_OP_all}E-H. Here $ \left\langle {\rho}_{a} \cdot {\rho}_{a} (\mathbf{r}) \right\rangle $ (red stripes) and $ \left\langle {\rho}_{b} \cdot {\rho}_{b} (\mathbf{r}) \right\rangle $ (blue stripes) represent the density-density correlation functions for charge modulations along \textbf{a} and \textbf{b}, respectively. The corresponding structure factors in momentum space $ {S}_{a} (\mathbf{Q})$ and $ {S}_{b} (\mathbf{Q})$ are represented as ellipsoids to reflect the presence of anisotropic correlations.

By inspecting the diagrams in Figs.\,S\ref{Stripe_CB_OP_all}E-H, one can recognize a common trait of checkerboard structures in momentum space, in that the following conditions have to be \textit{always} verified by symmetry:
\begin{align}
{\xi}^{a}_{x} &= {\xi}^{b}_{x}\nonumber\\
{\xi}^{a}_{y} &= {\xi}^{b}_{y},
\label{eq:CB_conditions_sup}
\end{align}
i.e., \textit{the correlation lengths along a given direction have to coincide for the two structure factors} ${S}_{a} (\mathbf{Q})$ \textit{and} ${S}_{b} (\mathbf{Q})$. At the bottom-left corner of the various diagrams we also report the conditions that apply to the correlation lengths in the various cases presented here (see also Table\,S\ref{Orders_constraints}). From this theoretical analysis, we can conclude that \textit{for both uniaxial and biaxial anisotropy it is in principle possible to discriminate between a pure checkerboard and a pure stripy order}, unless ${\xi}^{a}_{x}\!=\!{\xi}^{b}_{x}$, in which case the patterns shown in Figs.\,S\ref{Stripe_CB_OP_all}F and H become indistinguishable. Even though the uni- vs bi-directional character can be assessed regardless of the character of the anisotropy, the condition of biaxial anisotropy more intuitively discriminates between stripe and checkerboard charge order, as a consequence of a simple fact: for stripy nanodomains, the direction of the \textit{intra-domain} wavevector can be locked to the direction of anisotropy -- the two can be parallel (as in Fig.\,S\ref{Stripe_CB_OP_all}A) or perpendicular (not shown). Such configuration cannot be realized when the underlying order is checkerboard, since each single domain possesses both order parameters and therefore two wavevectors; this condition forbids by construction any locking to the axis of anisotropy. The case of domains which are oriented obliquely with respect to the crystallographic axes, but still giving rise to ellipsoidal structures in momentum space, can be equivalently treated by redefining the principal axes ($\mathbf{{x}^{\prime}}$, $\mathbf{{y}^{\prime}}$) for the domain structure. Ultimately, the same constraints that we derived for the correlation lengths along \textbf{x} and \textbf{y} would apply for $\mathbf{{x}^{\prime}}$ and $\mathbf{{y}^{\prime}}$ and the peak structure at $ {S}_{a} (\mathbf{Q})$ and $ {S}_{b} (\mathbf{Q})$ should still bare the same reciprocity conditions expressed in Eqs.\,\ref{eq:CB_conditions_sup} (but now referred to the $\mathbf{{x}^{\prime}}$ and $\mathbf{{y}^{\prime}}$ axes). This case is still not verified experimentally, and therefore rules out the presence of checkerboard order even in the more general case of charge order domains with different orientations. 

The comparison between the conditions laid out in Table\,S\ref{Orders_constraints} (second column) and the values for the correlation lengths determined experimentally (fourth column in Table\,S\ref{Orders_summary})
directly shows that the conditions for checkerboard order are violated for all YBCO samples -- indicating the presence of stripe order in the doping range surveyed in this study -- and at the same time allows classifying the type of charge order (see last column in Table\,S\ref{Orders_summary}). This classification indicates that Y651 and Y667 exhibit a biaxial stripy character, while Y675 can be categorized as having a uniaxial stripy character. This difference, as well as the fact that Y675 effectively displays a \textit{global} anisotropy with charge order domains preferentially elongated along a given direction, might relate to: (i) the crystal structure itself, with the increasing role of the orthorhombicity for increasing hole doping (the orthorhombic ratio $a/b$ increases monotonically); (ii) the proximity of the chain order peak, which in Y675 (Ortho-III) is located at $Q \!=\! 0.33$\,r.l.u.. Both aspects might lead to a stronger influence of the orthorhombicity and its intrinsically anisotropic, ${C}_{4}$-symmetry-broken structure onto the correlations and the domain structure in the CuO${}_{2}$ planes.
\\

\begin{figure}[t!]
\centering
\includegraphics[width=0.75\linewidth]{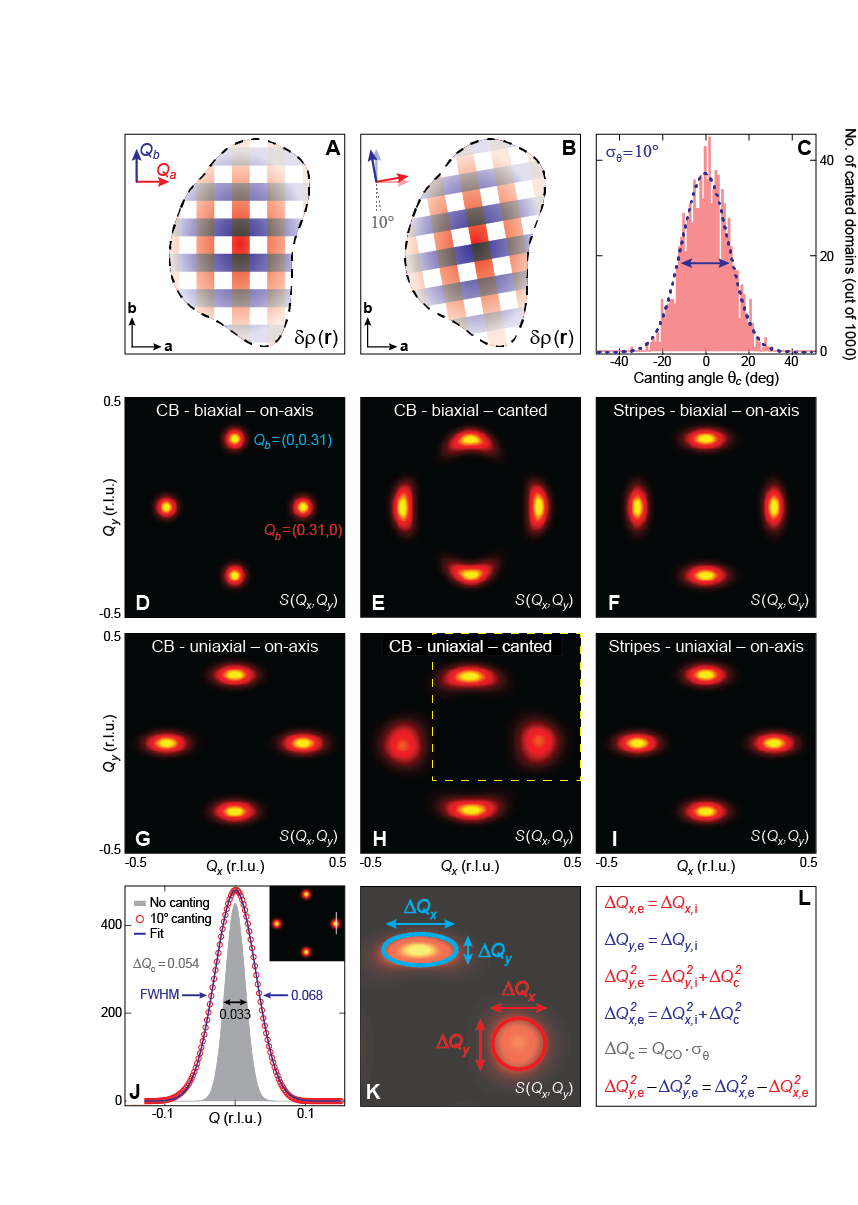}
\caption{\textbf{Domains with canted modulations and their effect on the CDW structure factor.} \textbf{(A,B)} On- and off-axis (${10}^{\circ}$ canting angle) checkerboard domain, respectively. \textbf{(C)} Distribution of canted domains, used in subsequent simulations. \textbf{(D-F)} Simulation of the structure factor $S \left( {Q}_{x}, {Q}_{y} \right)$ for a biaxial checkerboard-like charge order purely on-axis \textbf{(D)}, or with a Gaussian distribution of canted domains having $ \sigma \!=\! {10}^{\circ} $ \textbf{(E)}; and \textbf{(F)} for biaxial on-axis stripe-like order. \textbf{(G-I)} Same as \textbf{(D-F)}, but for uniaxial order. \textbf{(J)} Broadening induced by a distribution of canted domains with $ \sigma \!=\! {5}^{\circ} $, as seen via the linecut of $S \left( {Q}_{x}, {Q}_{y} \right)$ as indicated in the inset. \textbf{(K)} Definition of the peak widths for ${\mathbf{Q}}_{a}$ (red) and ${\mathbf{Q}}_{b}$ (blue), from the region highlighted in \textbf{(H)}. \textbf{(L)} Relations linking the experimental peak width ${\Delta Q}_{x/y,\mathrm{e}}$ to the intrinsic broadening ${\Delta Q}_{x/y,\mathrm{i}}$ and to the contribution due to the canted domains (${\Delta Q}_{\mathrm{c}}$) for the case of checkerboard order.}
\label{Canting_figure}
\end{figure}

\noindent {\bf Domain canting and its effect on the charge order structure in reciprocal space.} So far the charge order has been assumed to be strictly \textit{on-axis}, i.e. a modulation of the electronic density with crests (and valleys) parallel to the \textbf{a} or \textbf{b} axis for the ${\mathbf{Q}}_{b}$ and ${\mathbf{Q}}_{a}$ component, respectively. The existence of domains with slightly canted density modulations cannot be \textit{a priori} ruled out, hence we here analyze its effect on the momentum structure of charge order and its relationship to the peak shapes discussed in the previous section and their link to the inner structure of charge order (checkerboard vs. stripes). Figure\,S\ref{Canting_figure}A and S\ref{Canting_figure}B show the difference between on-axis and canted domains (with canting angle ${\theta}{\mathrm{c}} \!=\! {10}^{\circ}$ in Figure\,S\ref{Canting_figure}B).

Subsequently, we simulate the effect of a distribution of canted domains on the structure factor $ S \left({Q}_{x},{Q}_{x} \right) $. Such distribution is Gaussian and centered about ${\theta}_{\mathrm{c}} \!=\! {0}^{\circ}$ (see Fig.\,S\ref{Canting_figure}C, angular spread is ${\sigma}_{\theta} \!=\! {10}^{\circ}$); if otherwise, the charge order peaks would be displaced away from the reciprocal \textit{H} and \textit{K} axes, at variance with our experimental observations. Figure\,S\ref{Canting_figure}D shows the structure factor corresponding to a biaxial checkerboard structure with peak widths ${\Delta Q}_{x}^{a} \!=\! {\Delta Q}_{y}^{a} \!=\!{\Delta Q}_{x}^{b} \!=\!{\Delta Q}_{y}^{b} \!\simeq\! 0.033 $\,r.l.u. (see Fig.\,S\ref{Canting_figure}K for a graphical definition), in the absence of canted domains. The effect of a distribution like that of Fig.\,S\ref{Canting_figure}C is showcased in Fig.\,S\ref{Canting_figure}E, while Fig.\,S\ref{Canting_figure}F shows a simple (no canting) biaxial stripe scenario with ${\Delta Q}_{x}^{a} \!=\! {\Delta Q}_{y}^{b} \!=\! 0.033$ and ${\Delta Q}_{x}^{b} \!=\!{\Delta Q}_{y}^{b} \!=\! 0.066 $\,r.l.u. (Figs\,S\ref{Canting_figure}G-I show the same comparison for the case of unixial order, which also implies anisotropic correlation lengths and peak shape). The effect of the canting angle distribution is immediately evident in that it broadens the original charge order peaks in the transverse direction, as expected since canting displaces the peak along a circle centered at $Q \!=\! 0$ and with radius $Q \!=\! {Q}_{\mathrm{CO}}$. Also, for the same reason, the peaks tend to get curved at their edges, an effect which becomes particularly evident when ${\sigma}_{\theta} \!\gtrsim\! {5}^{\circ}$.

A qualitative analysis reveals a striking similarity between panels E and F, thus raising the possibility that, contrary to the conclusions reached in the previous section, the pattern shown in Figs.\,S\ref{Canting_figure} might in fact be compatible with checkerboard, when a distribution of canted domains is present. However, a quantitative analysis of the experimental peak widths shows that a checkerboard scenario has to be still ruled out. In the following, we consider the more general checkerboard case, for which these relations must hold (see also Supplementary Table\,S\ref{Orders_constraints}):
\begin{equation}
{\Delta Q}^{a}_{x,\mathrm{i}} \!=\! {\Delta Q}^{b}_{x,\mathrm{i}} \lessgtr {\Delta Q}^{b}_{y,\mathrm{i}} \!=\! {\Delta Q}^{a}_{y,\mathrm{i}}
\label{eq:CB_constraints}
\end{equation}
The subscript \textit{i} means \textit{intrinsic}, as the above relations refer to the native broadening of the charge order peaks resulting from finite spatial correlations. As mentioned earlier, the effect of a distribution of canted domains will change the transverse width. The momentum broadening associated to the distribution of canted domains (${\Delta Q}_{\mathrm{c}}$), in the case of small angular spreads, can be approximated as ${\Delta Q}_{\mathrm{c}} \!=\! 2 {Q}_{\mathrm{CO}} \cdot {\sigma}_{\theta}$ which, for ${\sigma}_{\theta} \!=\! {5}^{\circ} \!=\! 0.087$\,rad, gives ${\Delta Q}_{\mathrm{c}} \!\simeq\! 0.054$\,r.l.u.. The analysis of Fig.\,S\ref{Canting_figure}J demonstrates that the extra broadening adds up in quadrature to the intrinsic one (${0.068}^{2} \!\sim\! {0.054}^{2} + {0.033}^{2}$), by comparing the FWHM of the CO peaks along the transverse direction (see inset) before ($0.033$\,r.l.u.) and after ($0.068$\,r.l.u.) the inclusion of a domain distribution with ${\sigma}_{\theta} \!=\! {5}^{\circ}$.

As a result, while the experimental longitudinal widths will remain unaffected:
\begin{equation}
{\Delta Q}^{a}_{x,\mathrm{e}} \!=\! {\Delta Q}^{a}_{x,\mathrm{i}} \qquad {\Delta Q}^{b}_{y,\mathrm{e}} \!=\! {\Delta Q}^{b}_{y,\mathrm{i}}
\label{eq:Long_broadening_canting}
\end{equation}
the transverse widths will increase, and experimentally one would observe: 
\begin{eqnarray}
{\Delta Q}^{a}_{y,\mathrm{e}} = \sqrt{{\left( {\Delta Q}^{a}_{y,\mathrm{i}} \right)}^{2} + {\left( {\Delta Q}_{\mathrm{c}} \right)}^{2}} = \sqrt{{\left( {\Delta Q}^{b}_{y,\mathrm{i}} \right)}^{2} + {\left( {\Delta Q}_{\mathrm{c}} \right)}^{2}} = \sqrt{{\left( {\Delta Q}^{b}_{y,\mathrm{e}} \right)}^{2} + {\left( {\Delta Q}_{\mathrm{c}} \right)}^{2}}\nonumber\\
{\Delta Q}^{b}_{x,\mathrm{e}} = \sqrt{{\left( {\Delta Q}^{b}_{x,\mathrm{i}} \right)}^{2} + {\left( {\Delta Q}_{\mathrm{c}} \right)}^{2}} = \sqrt{{\left( {\Delta Q}^{a}_{x,\mathrm{i}} \right)}^{2} + {\left( {\Delta Q}_{\mathrm{c}} \right)}^{2}} = \sqrt{{\left( {\Delta Q}^{a}_{x,\mathrm{e}} \right)}^{2} + {\left( {\Delta Q}_{\mathrm{c}} \right)}^{2}}
\label{eq:Transverse_broadening_canting}
\end{eqnarray}
where in the second (third) equality Eqs.\,\ref{eq:CB_constraints} (Eqs.\,\ref{eq:Long_broadening_canting}) have been used. By taking the square of Eqs.\,\ref{eq:Transverse_broadening_canting} it directly follows that, \textit{in the case of checkerboard order in presence of a distribution of canted domains centered about the crystallographic axes}, the following equality must hold:
\begin{equation}
{\left( {\Delta Q}^{a}_{y,\mathrm{e}} \right)}^{2} - {\left( {\Delta Q}^{b}_{y,\mathrm{e}} \right)}^{2} = {\left( {\Delta Q}^{b}_{x,\mathrm{e}} \right)}^{2} - {\left( {\Delta Q}^{a}_{x,\mathrm{e}} \right)}^{2}
\label{eq:CB_canting_relation}
\end{equation}
The set of mathematical relations introduced so far has also been summarized in Fig.\,S\ref{Canting_figure}L.

In addition, Eq.\,\ref{eq:CB_canting_relation} has been evaluated for the three YBCO samples, and the results are reported in Supplementary Table\,S\ref{CB_canting_condition}, showing that such relation is always violated. This leads to the conclusion that, even in the presence of a distribution of domains with canted charge modulations, the checkerboard scenario remains incompatible with the experimental results.
\\

\begin{figure}[t!]
\centering
\includegraphics[width=1\linewidth]{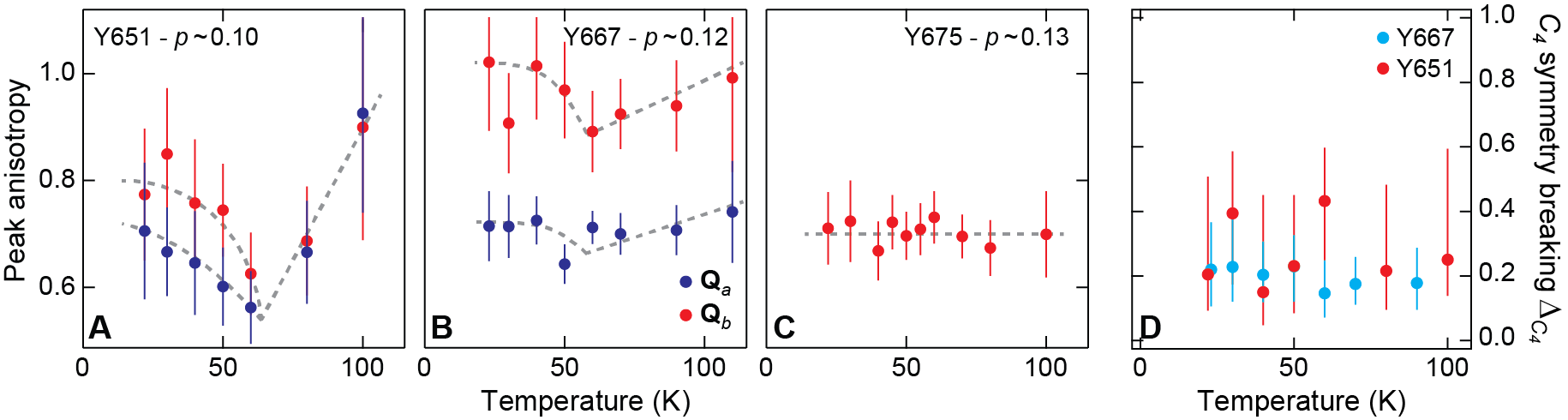}
\caption{\textbf{CDW peak anisotropy and ${C}_{4}$ symmetry breaking vs. temperature.} \textbf{(A-C)} Peak anisotropy ${\gamma}_{a} \!=\! {\Delta Q}_{a,\parallel} / {\Delta Q}_{a,\perp}$ (blue markers) and ${\gamma}_{b} \!=\! {\Delta Q}_{b,\parallel} / {\Delta Q}_{b,\perp}$ (red) for Y651, Y667, and Y675, respectively. The uncertainties (error bars) on line width data points $\Delta Q$ have been estimated from the fitting analysis of RXS scans using standard procedures [48], and subsequently converted to uncertainties for the anisotropy $\gamma$ using error propagation. Grey dashed lines are guides-to-the-eye. \textbf{(D)} ${C}_{4}$ symmetry breaking order parameter ${\Delta}_{{C}_{4}}$ as a function of temperature, for Y651 (red) and Y667 (light blue), respectively. Uncertainties (error bars) on the values of ${\Delta}_{{C}_{4}}$ have been estimated by taking the maximum margin after having evaluated ${\Delta}_{{C}_{4}}$ at the extrema of the error range, i.e. for all combinations of: $ \lbrace {\Delta Q}_{a,\parallel} \pm {\delta}_{{\Delta Q}_{a,\parallel}}; {\Delta Q}_{a,\perp} \pm {\delta}_{{\Delta Q}_{a,\perp}}; {\Delta Q}_{b,\parallel} \pm {\delta}_{{\Delta Q}_{b,\parallel}}; {\Delta Q}_{b,\perp} \pm {\delta}_{{\Delta Q}_{b,\perp}} \rbrace$.}
\label{Aniso_stripyness_Tdep}
\end{figure}

\noindent {\bf Doping- and temperature-dependent CDW peak asymmetry vs. YBCO orthorhombicity.} In this last section, we focus on a more detailed analysis of the CDW peak asymmetry. In this case, the available experimental data allow extracting the longitudinal and transverse correlation lengths for the YBCO dopings and CDW peaks reported in Fig.\,3 of the main text. Here we extend such analysis to the extraction of the peak anisotropy defined as $\gamma \!=\! {\Delta Q}_{\parallel} / {\Delta Q}_{\perp}$ as a function of temperature, in a range where the peak widths can be extracted with reasonable precision ($T \!<\! 100$\,K). The results are shown in Fig.\,S\ref{Aniso_stripyness_Tdep}A-C and suggest a non-monotonous temperature dependence of ${\gamma}_{a,b}$, which is more evident in Y651 and Y667. In particular, the peak anisotropy increases upon cooling down towards $T \!>\! {T}_{\mathrm{c}}$, then is maximized around ${T}_{\mathrm{c}}$, and eventually recovers below ${T}_{\mathrm{c}}$, consistent with the pronounced directionality of the superconductivity-induced suppression of CDW correlations (see again Fig.\,3 in the main text).

The magnitude of the asymmetry, which is closely related to the degree of elongation of the CDW peaks in momentum space (see again Fig.\,1 in the main text), allows drawing important considerations with respect to the role of orthorhombicity in our study. The following points will clarify a few key aspects of our data that directly rule out the possibility that the observed structures are solely driven by the structural symmetry breaking associated to the chain layer:

\begin{itemize}
\item In general, in presence of isotropic correlations (measured in number of unit cells) defined on a square lattice, the two-dimensional structure factor is also isotropic. However, when the underlying structure and its unit cell become orthorhombic, the peaks will develop an elongation with an associated anisotropy $\gamma$ equivalent to the orthorhombic ratio $a/b$ (or $b/a$), which is equal to $a/b \!=\! 0.9902$ ($b/a \!=\! 1.0098$) in Y651, $a/b \!=\! 0.9856$ ($b/a \!=\! 1.0145$) in Y667, and $a/b \!=\! 0.9845$ ($b/a \!=\! 1.0157$) in Y675. The experimental values shown in Fig.\,S\ref{Aniso_stripyness_Tdep}A-C in almost all cases strongly deviate from the value found above for the orthorhombic ratio, indicating a much larger anisotropy. Furthermore, the doping dependence of the anisotropy, which switches from biaxial to uniaxial across $p=0.12$, is also inconsistent with a behavior completely driven by the structural orthorhombicity  (we note again that our samples are fully detwinned).
\item If we assume that the influence acted by the chain layer and its associated potential onto the CuO${}_{2}$ plane is more complex and, e.g., involves native anisotropic correlations in the CuO chain order, than one might wonder whether this alone could explain the observed anisotropy in the peak profiles. If this was the case, then the CDW peak along $\mathbf{a}$ and $\mathbf{b}$ should exhibit the very same elongation, since the chain layer breaks ${C}_{4}$ in a unique manner. In other words, the anisotropy measured at ${\mathbf{Q}}_{a}$ and ${\mathbf{Q}}_{b}$ must be the same within such scenario, again at variance with the experimental data presented in Fig.\,S\ref{Aniso_stripyness_Tdep}A-C. 
\item The deviation of the peak anisotropy parameter $\gamma$ from the isotropic case ($\gamma \!=\!1$) is largest for the most underdoped sample, and is reduced for increasing hole doping (see Fig.\,S\ref{Symmetry_OP_rot}B), in stark contrast with the evolution of the crystal orthorhombicity which instead grows as optimal doping is approached.
\item The temperature evolution of the peak anisotropy shows an anomaly near the superconducting transition ${T}_{\mathrm{c}}$. Such a behavior in principle can not be immediately deemed to be incompatible with the evolution of the lattice parameters $a$ and $b$, whose temperature-dependent espansivity ${\alpha}_{a,b}$ also shows a cusp near ${T}_{\mathrm{c}}$ [36]. However, the anomaly associated to the crystal orthorhombicity [36] gets more and more pronounced for increasing doping (with a maximum at optimal doping reflecting the complete oxygenation of the chains), and is therefore inconsistent with our finding of a charge order peak anisotropy that is more enhanced in the underdoped region, and weakens above $p\!=\!0.12$.
\item In addition to these aspects, a recent study by \textit{Achkar et al.} [47] directly showed that the CDW correlations are to a large degree unaffected upon disordering the CuO chains, and therefore are intrinsic to the CuO${}_{2}$ planes. This once again validates the scenario where the ordering mechanisms in the planes are largely insensitive to the phenomenology in the chain layer, where the orthorhombicity originates from.
\end{itemize}

Finally, Figure\,S\ref{Aniso_stripyness_Tdep}D shows the ${C}_{4}$ symmetry breaking parameter ${\Delta}_{{C}_{4}}$ as a function of temperature below 100\,K, for Y651 and Y667 (in Y675 no T-dependent data for ${\mathbf{Q}}_{a}$ are available, and therefore ${\Delta}_{{C}_{4}}$ cannot be evaluated). The amount of ${C}_{4}$ symmetry breaking ranges around 0.2 for both doping levels, although both the error bars and the data scatter are larger for Y651, due to the generally lower CDW peak intensity. In any case, we can conclude that ${\Delta}_{{C}_{4}}$ does not show a pronounced temperature dependence, which seems to suggest that while the overall amplitude of density correlations in the CuO${}_{2}$ planes changes with temperature, their stripe-like character does not instead change significantly.

\clearpage

\subsection*{Supplementary Figures}

\clearpage

\subsection*{Supplementary Tables}

\begin{table}[h!]
\begin{center}
\begin{tabular}{ccc}
\toprule
\toprule
Order type & Conditions on correlations & Conditions on peak width \\
\midrule
\midrule
Stripy biaxial & ${\xi}^{a}_{x} \gtrless {\xi}^{a}_{y}$; ${\xi}^{b}_{x} \lessgtr {\xi}^{b}_{y}$ & ${\Delta Q}^{a}_{x} \lessgtr {\Delta Q}^{a}_{y}$; ${\Delta Q}^{b}_{x} \gtrless {\Delta Q}^{b}_{y}$ \\
\midrule
Stripy uniaxial & ${\xi}^{a}_{x} \gtrless {\xi}^{a}_{y}$; ${\xi}^{b}_{x} \gtrless {\xi}^{b}_{y}$ & ${\Delta Q}^{a}_{x} \gtrless {\Delta Q}^{a}_{y}$; ${\Delta Q}^{b}_{x} \gtrless {\Delta Q}^{b}_{y}$ \\
\midrule
Checkerboard biaxial & ${\xi}^{a}_{x} \!=\! {\xi}^{a}_{y} \!=\! {\xi}^{b}_{x} \!=\! {\xi}^{b}_{y}$ & ${\Delta Q}^{a}_{x} \!=\! {\Delta Q}^{a}_{y} \!=\! {\Delta Q}^{b}_{x} \!=\! {\Delta Q}^{b}_{y}$ \\
\midrule
Checkerboard uniaxial & ${\xi}^{a}_{x} \!=\! {\xi}^{b}_{x} \gtrless {\xi}^{b}_{y} \!=\! {\xi}^{a}_{y}$ & ${\Delta Q}^{a}_{x} \!=\! {\Delta Q}^{b}_{x} \lessgtr {\Delta Q}^{b}_{y} \!=\! {\Delta Q}^{a}_{y}$ \\
\bottomrule
\bottomrule
\end{tabular}
\end{center}
\caption{{\bf Constraints on correlation lengths and peak elongations for checkerboard and stripe orders}. Reported in this table are the conditions that restrain the correlation lengths between the two directions and charge order peaks based on general considerations.}
\label{Orders_constraints}
\end{table}
\begin{table}[h!]
\begin{center}
\begin{tabularx}{1\textwidth}{ccc *{2}{Y}}
\toprule
\toprule
Sample & Doping & Wavevector & Correlations ($\AA$) & Order type \\
\midrule
\midrule
Y651 & \multirow{2}{*}{$\sim\! 0.10$} & ${\mathbf{Q}}_{a}$ & ${\xi}_{x}\!=\!43 (\pm 7) > {\xi}_{y}\!=\!33 (\pm 1)$ & \multirow{2}{*}{Stripy biaxial} \\
Y651 & & ${\mathbf{Q}}_{b}$ & ${\xi}_{x}\!=\!27 (\pm 1) < {\xi}_{y}\!=\!41 (\pm 2)$ & \\
\midrule
Y667 & \multirow{2}{*}{$\sim\! 0.12$} & ${\mathbf{Q}}_{a}$ & ${\xi}_{x}\!=\!45 (\pm 2) > {\xi}_{y}\!=\!42 (\pm 2)$ & \multirow{2}{*}{Stripy biaxial} \\
Y667 & & ${\mathbf{Q}}_{b}$ & ${\xi}_{x}\!=\!35 (\pm 1) < {\xi}_{y}\!=\!51 (\pm 3)$ & \\
\midrule
Y675 & \multirow{2}{*}{$\sim\! 0.13$} & ${\mathbf{Q}}_{a}$ & ${\xi}_{x}\!=\!47 (\pm 4) < {\xi}_{y}\!=\!55 (\pm 4)$ & \multirow{2}{*}{Stripy uniaxial} \\
Y675 & & ${\mathbf{Q}}_{b}$ & ${\xi}_{x}\!=\!38 (\pm 1) < {\xi}_{y}\!=\!54 (\pm 2)$ & \\
\bottomrule
\bottomrule
\end{tabularx}
\end{center}
\caption{{\bf Charge order classification in underdoped YBCO compounds}. Reported in this table are the correlation lengths as extracted from the RXS data. Comparison with the momentum structure of the order parameter allows classifying the different types of order (rightmost column).}
\label{Orders_summary}
\end{table}
\begin{table}[h!]
\begin{center}
\begin{tabular}{ccc}
\toprule
\toprule
Sample & ${\left( {\Delta Q}^{a}_{y,\mathrm{e}} \right)}^{2} - {\left( {\Delta Q}^{b}_{y,\mathrm{e}} \right)}^{2}$ & ${\left( {\Delta Q}^{b}_{x,\mathrm{e}} \right)}^{2} - {\left( {\Delta Q}^{a}_{x,\mathrm{e}} \right)}^{2}$ \\
\midrule
\midrule
Y651 & $32\,(\pm 25) \times {10}^{-5}\,{\AA}^{-2}$ & $77\,(\pm 13) \times {10}^{-5}\,{\AA}^{-2}$ \\
\midrule
Y667 & $18\,(\pm 9) \times {10}^{-5}\,{\AA}^{-2}$ & $36\,(\pm 9) \times {10}^{-5}\,{\AA}^{-2}$ \\
\midrule
Y675 & $-4\,(\pm 9) \times {10}^{-5}\,{\AA}^{-2}$ & $23\,(\pm 9) \times {10}^{-5}\,{\AA}^{-2}$ \\
\bottomrule
\bottomrule
\end{tabular}
\end{center}
\caption{{\bf Evaluation of condition for checkerboard order in the presence of canted domains.} In order for a checkerboard+canted domains scenario to be compatible with experimental data, the numbers on the second and third column must be the same. When these expressions are evaluated and compared with the respective error bars, we find that the equality is violated for all three YBCO compounds.}
\label{CB_canting_condition}
\end{table}

\clearpage

\subsection*{References and Notes}

[44] J. D. Jorgensen, B. Veal, A. Paulikas, L. Nowicki, G. Crabtree, H. Claus, W. Kwok, Structural properties of oxygen-deficient YBa${}_{2}$Cu${}_{3}$O${}_{7-\delta}$. \textit{Phys. Rev. B} \textbf{41}, 1863-1877 (1990).\\

\noindent
[45] R. Liang, D. A. Bonn, W. N. Hardy, Evaluation of CuO${}_{}2$ plane hole doping in YBa${}_{2}$Cu${}_{3}$O${}_{6+x}$ single crystals. \textit{Phys. Rev. B} \textbf{73}, 180505 (2006).\\

\noindent
[46] K. Takubo, R. Comin, D. Ootsuki, T. Mizokawa, H. Wadati, Y. Takahashi, G. Shibata, A. Fujimori, R. Sutarto, F. He, S. Pyon, K. Kudo, M. Nohara, G. Levy, I. S. Elfimov, G. A. Sawatzky, A. Damascelli, Bond order and the role of ligand states in stripe-modulated IrTe${}_{2}$. \textit{Phys. Rev. B} \textbf{90}, 081104 (2014).\\

\noindent
[47] A. J. Achkar, X. Mao, C. McMahon, R. Sutarto, F. He, R. Liang, D. A. Bonn, W. N. Hardy, D. G. Hawthorn, Impact of quenched oxygen disorder on charge density wave order in YBa${}_{2}$Cu${}_{3}$O${}_{6+x}$. \textit{Phys. Rev. Lett.} \textbf{113}, 107002 (2014).\\

\noindent
[48] The data-fitting routines used for this analysis implicitly assume that the errors are normally
distributed with zero mean and constant variance and that the fit function is a good
description of the data. The coefficients and their sigma values (representing uncertainties)
are estimates of what one would get if the same fit were performed an infinite number of
times on the same underlying data (but with different noise each time) and then the mean
and standard deviation were calculated for each coefficient


\begin{thebibliography}{10}
\bibitem{Charge_note}
Because of our inability to measure magnetic ordering, our concept of stripes
  only applies to the charge modulations, unlike the canonical definition used
  in La-based cuprates, which also implies concomitant magnetic order.
\bibitem{DeltaQ_note}
Unless ${\Delta Q}^{a}_{x}\!=\!{\Delta Q}^{b}_{x}$, in which case the patterns
  shown in Figs.\,\ref{Fig2b}, F and H become indistinguishable. This
  situation, however, is not encountered in our study.
\end{thebibliography}
\end{document}